\documentclass[aps,prd,showpacs,preprintnumbers,nofootinbib,amsmath,amssymb]{revtex4}

\usepackage{graphicx}
\usepackage{dcolumn}
\usepackage{bm}
\usepackage{amsmath}

\newcommand{\tla}{\tilde\lambda}
\newcommand{\ie}{{\it i.e.\ }}
\newcommand{\eg}{{\it e.g.\ }}

\begin{document}

\rightline{}
\title{The Maximal Abelian Gauge 
in $SU(N)$ gauge theories and thermal monopoles for $N = 3$.}

\author{Claudio Bonati}
\email{bonati@df.unipi.it}
\affiliation{
Dipartimento di Fisica dell'Universit\`a
di Pisa and INFN - Sezione di Pisa,\\ Largo Pontecorvo 3, I-56127 Pisa, Italy}

\author{Massimo D'Elia}
\email{delia@df.unipi.it}
\affiliation{
Dipartimento di Fisica dell'Universit\`a
di Pisa and INFN - Sezione di Pisa,\\ Largo Pontecorvo 3, I-56127 Pisa, Italy}

\date{\today}

\begin{abstract}
We discuss and propose a proper extension of the Abelian projection based
on the Maximal Abelian Gauge to $SU(N)$ gauge theories.
Based, on that, we investigate the properties of thermal 
Abelian monopoles in the deconfined phase of the 
$SU(3)$ pure gauge theory. Such properties are very similar
to those already found for $SU(2)$, confirming the relevance
of the magnetic component close to $T_c$ and the 
possible condensation of thermal monopoles as 
the deconfinement temperature is crossed from above. Moreover, 
we study the correlation functions among monopoles related to different
$U(1)$ subgroups, which show interesting features and reveal 
the presence of non-trivial interactions.
\end{abstract}


\maketitle

\section{Introduction}

An exact identification of the mechanism 
responsible for Color Confinement, and of the 
effective degrees of freedom relevant to it, is still missing.
A commonly accepted scenario is that such degrees of freedom must be of 
dual, topological nature. 
A possible proposed mechanism 
is that based on dual superconductivity~\cite{thooft75,mandelstam},
i.e. on the idea that the QCD vacuum is characterized by the
spontaneous breaking of a magnetic symmetry, induced by the condensation
of magnetically charged defects.

Possible approaches, followed to test this mechanism by lattice
simulations, have either 
checked for the spontaneous symmetry breaking, by looking at
the vacuum expectation value of magnetically charged operators
and at the effective action of monopole 
configurations~\cite{super0,superI-II,superIII,superfull,superIV,moscow,bari,superlast}, or have looked at the properties of monopole configurations extracted
from non-Abelian configurations generated at equilibrium.
The identification of Abelian degrees of freedom in 
non-Abelian gauge theories, 
and of Abelian 
monopoles in particular, relies on a 
procedure known as Abelian projection, which is based 
on the choice of an adjoint
field. Since no natural adjoint field exists in QCD, that implies some
arbitrariness. A popular choice is to perform the projection in 
the so-called Maximal Abelian gauge (MAG).

It is natural to expect that topological degrees of freedom, 
which are responsible for color confinement,
play a relevant role also at and around the deconfinement transition.
Indeed, magnetic monopoles evaporating from the low temperature magnetic 
condensate, which are usually known as thermal monopoles,
have been advocated for their possible role in the properties
of strongly interacting matter above the deconfinement 
transition~\cite{shuryak,chezak}
Thermal monopoles are identified, in lattice QCD simulations at finite 
temperature, by looking for  
monopole currents with non-trivial wrappings in the Euclidean 
temporal direction~\cite{chezak,bornya92,ejiri}.  
Systematic lattice studies, regarding the properties of thermal monopoles in  
the deconfined phase of $SU(2)$ Yang-Mills theory have been performed
in Refs.~\cite{monden,moncon,dyons,born1,born2,born3,braguta}.

Results for the $SU(2)$ gauge theory have shown several interesting 
properties of such objects, which are compatible with a scenario
in which the magnetic component of the Quark-Gluon Plasma
plays a significant role right above the deconfinement temperature,
while it is less relevant at asymptotically high temperatures,
where the plasma is electrically dominated~\cite{liao,ratti}. 
In particular, spatial correlations of thermal monopoles show
the presence of Coulomb-like, screened interactions
among monopoles and antimonopoles~\cite{monden}, 
with an effective magnetic coupling 
which grows in the high $T$ regime~\cite{liao}, where
the density of monopoles is also logarithmically suppressed
with respect to the electric component. 
On the contrary, the coupling 
decreases approaching the low $T$ regime, and
the analysis of the statistical distribution 
of trajectories with multiple wrappings in the temporal
direction~\cite{cristof,moncon} suggests that thermal monopoles may condense
at a temperature which coincides, within errors,
with the deconfinement temperature, giving further support
to a confinement mechanism based on the 
condensation of magnetic charges.

The purpose of the present study is to extend such investigation
to the pure gauge theory with three colors; 
preliminary results in the same direction have been 
reported in Ref.~\cite{bornsu3} and will be discussed in the following.
The maximal Abelian
subgroup of $SU(N)$ gauge theories is $U(1)^{(N-1)}$, hence
the main change, when going to $N > 2$, is that various Abelian charges, i.e.
various different species of monopoles, can be identified.
On one hand, that makes a proper extension of the definition
of the Maximal Abelian Gauge less trivial. On the other hand, 
new properties appear, associated with the interactions
among different monopole species.

The paper is organized as follows. 
The extension of MAG to generic $SU(N)$ gauge groups  will be discussed
in detail in Section~\ref{magsun}. Numerical results obtained for 
thermal monopoles in the deconfined phase of $SU(3)$ will be 
presented and discussed in Section~\ref{monden}. Finally,
in Section~\ref{discon}, we draw our conclusions. Technical 
details and comparison among different definitions of MAG
are reported in Appendixes~\ref{magalg}, \ref{takediag} and 
\ref{tlachange}.

\section{Abelian projection and MAG monopoles in $SU(N)$}
\label{magsun}

In the following we shall review the definition of Abelian projection
and of abelian magnetic monopoles in $SU(N)$ gauge theories. 
Then we shall focus on the Abelian projection defined by the 
so-called Maximal Abelian Gauge, discussing how the standard
$SU(2)$ definition can be properly extended to $N > 2$. 
Even if some of the facts reported in this Section are already 
known from the literature, we report them here for the 
reader's convenience.

\subsection{Abelian projection and monopoles}
\label{2A}

Abelian projection is the procedure for identifying Abelian $U(1)$ gauge 
symmetries within a non-Abelian theory. In the case of a $SU(N)$ gauge group, starting from 
a generic local field $\phi (x) = \sum_a \phi^a(x) T^a$ (with $\sum_a\phi^a\phi^a=$ const.)
transforming in the adjoint representation of the gauge group, one can define 
the Abelian 't Hooft tensor \cite{thooft74}:
\begin{equation}
F_{\mu\nu} = {\rm tr} \left( \phi\ G_{\mu\nu} \right) - 
\frac{i}{g} {\rm tr} \left( \phi \left[ D_\mu \phi, D_\nu \phi  
\right] \right)
\label{thoofttensor}
\end{equation}
where, as usual, 
\begin{equation*}
G_{\mu\nu} = \partial_\mu A_\nu - \partial_\nu A_\mu + i g
\left[ A_\mu,A_\nu\right] \,\, ; \,\,\,\,\,\, 
A_\mu = A_\mu^a T^a \,\, ; \,\,\,\,\,\, 
D_\mu\phi = \partial_\mu\phi - ig \left[A_\mu,\phi\right]
\end{equation*}
and the $SU(N)$ generators are normalized as 
${\rm tr} (T^a T^b) = \delta^{ab}/2$. The 't Hooft tensor is, by construction,
a gauge invariant quantity, which however depends on the choice of the 
adjoint field $\phi$.

One can prove (see Refs.~\cite{sun_proj, dlp} for a detailed discussion) that 
terms bilinear in $A_\mu$ and terms containing $A_\mu \partial_\nu \phi$ cancel in 
Eq.~(\ref{thoofttensor}) in gauges where $\phi$ is a constant diagonal field, 
$\phi(x) = \Phi_D$, and if $\Phi_D$ is one of the $N - 1$ fields
\begin{equation}
\phi_{0}^{k}=\frac{1}{N}\, {\rm diag}\, (\, \underbrace{N-k,\ldots,N-k}_{k}\, , 
\underbrace{ - k ,\,\ldots , - k }_{N-k})
\label{fwdef}
\end{equation} 
where $k = 1\ldots N-1$, corresponding to the fundamental weights of the $SU(N)$ 
algebra. In such gauge (usually known as {\em unitary gauge})
$F_{\mu \nu}$ reduces to a standard electromagnetic tensor,  
\begin{equation}
F^{(k)}_{\mu\nu} = {\rm tr} \left( \partial_\mu (\phi_0^k A_\nu) - 
\partial_\nu ( \phi_0^k A_\mu) \right)
\equiv \partial_\mu a^{(k)}_\nu - 
\partial_\nu a^{(k)}_\mu\, ,
\label{thoofttensor_abelian}
\end{equation} 
where we have defined 
\begin{equation}
a^{(k)}_\mu \equiv {\rm tr} (\phi_0^k A_\mu) = \sum_{j = 1}^k ({A_\mu})_{jj} \, ,
\label{amudef}
\end{equation}
as can be verified by exploiting the fact that $A_\mu$ is traceless.
If $A^D_\mu$ is the diagonal part of the gauge field $A^a_\mu T^a$,
then it is trivial to check that 
\begin{equation*}
A^D_\mu = \sum_{k=1}^{N-1} a^{(k)}_\mu \alpha^k
\end{equation*}
where 
\begin{equation*}
\alpha^k = \frac{1}{2} 
{\rm diag }\,(0,0,\ldots,0,\overbrace{1,-1}^{k,k+1},0,\ldots,0)
\end{equation*}
is the matrix associated with the $k$-th simple root. Hence 
the 't Hooft tensor $F^{(k)}_{\mu\nu}$ is the electromagnetic 
tensor associated to the $U(1)$ subgroup generated, in the diagonal
gauge, by $\alpha^k$.

In $SU(N)$ gauge theories, it is possible to identify $N-1$ independent
$U(1)$ subgroups, \ie a maximal $U(1)^{(N-1)}$ Abelian subgroup. 
The standard procedure~\cite{thooft81} is to identify a local, 
hermitian operator $X(x)$ transforming covariantly in the adjoint 
representation, $X(x) \to G(x) X(x) G^{-1}(x)$, then fixing the gauge 
so as to make $X(x)$ diagonal everywhere 
\begin{equation}
X(x) = X^D(x) = {\rm diag} (X_1(x), X_2(x), \dots , X_N(x)) \, .
\end{equation} 
That indeed fixes the gauge apart from a residual $U(1)^{(N-1)}$ gauge 
symmetry. This is actually true only if a given ordering is also assigned for 
the eigenvalues $X_i(x)$, otherwise the residual invariance group contains 
also permutations. Since the eigenvalues are real, a standard possible choice 
\cite{thooft81} is 
\begin{equation}
X_j(x) \geq X_{j+1}(x) \, .
\label{ordering}
\end{equation}
A 't Hooft tensor $F^{(k)}_{\mu\nu}$ can then be associated to each of the residual
$U(1)$ group, fixing $\phi(x) = \phi_0^k$ in the diagonal gauge.
$X(x)$ can be chosen, without any loss of generality, to be a traceless operator, 
then $X^D(x)$ can be written in terms of the $\phi_0^k$ matrices, which form a 
complete set over the Cartan subalgebra, as follows:
\begin{equation}
X^D(x) = \sum_{k = 1}^{N-1} c^k(x) \phi_0^k\, , 
\label{xdexpansion}
\end{equation}
and making use of the relation ${\rm tr} (\alpha^k \phi_0^{k'}) = \delta^{k k'}/2$ 
we have 
\begin{equation*}
c^k(x) = \frac{1}{2} {\rm tr} (\alpha^k X^D(x)) = X_k(x) - X_{k+1}(x)\ .
\end{equation*}

A special role is played by points where one of the coefficients
in Eq.~(\ref{xdexpansion}) vanishes, \ie where two consecutive 
eigenvalues of $X(x)$ coincide. Suppose $c^k(x_0) = 0$ for some $k$, \ie 
$X_k(x) = X_{k+1}(x)$, then in $x_0$ the residual $U(1)$ symmetry, corresponding
to gauge transformations $G(x) = \exp(i \omega \alpha^k)$,
is  enlarged to the full $SU(2)$ invariance subgroup associated with the simple root $\alpha^k$. 
For that reason, the point $x_0$ represents a topological defect, where the gauge fixing 
procedure is not well defined. The projection of the operator $X(x)$ over the $k$-th 
$SU(2)$ subgroup vanishes in $x_0$; around $x_0$, instead, one can choose either 
a hedgehog solution for it \cite{thooft74, poly74} or, in the unitary gauge, a solution 
where $X(x)$ is diagonal and the field $a^{(k)}_\mu$ contains the contribution 
from a Coulomb-like magnetic field centered around $x_0$. 
In both cases we identify $x_0$ as the location of a magnetic monopole, relatively
to the corresponding $U(1)$ subgroup, where the Abelian and the non-Abelian 
Bianchi identities are violated~\cite{nabianchi}. 

In the particular case of $SU(3)$, which is the subject of our numerical
study, one can identify two different Abelian monopole species which
manifest themselves, in the unitary gauge, by the presence 
of a Coulomb-like magnetic field in the corresponding Abelian gauge fields
\begin{equation}
a^{(1)}_\mu  = ({A_\mu})_{11} \,\,\, ; \,\,\,\,\,\,\,\,\,\,\,\,
a^{(2)}_\mu  = ({A_\mu})_{11} + ({A_\mu})_{22} = - ({A_\mu})_{33}
\, . 
\label{defamusu3}
\end{equation}

\subsection{Abelian projection and monopoles on the lattice}
\label{2B}

Let us now discuss how that applies to the lattice formulation of
non-Abelian gauge theories, a topic first studied in Ref.~\cite{mag3_0}. 
The Abelian projection is defined, as in the 
continuum, in terms of a local operator $X(n)$, living on lattice sites $n$,
transforming in the adjoint representation. Differently from the continuum
case, however, ambiguities emerge when defining the 't~Hooft tensor
and the Abelian gauge fields in the unitary gauge, where 
$X(n)$ is diagonal. Gauge links $U_\mu (n)$, which are 
the elementary lattice variables, are related to the gauge field 
$A_\mu$ by the following relation, 
$U_\mu(n) \simeq_{a \to 0} \exp (i a g A_\mu(n)) \simeq_{a \to 0} 
1 + i a g A_\mu(n)$,
which is valid in the continuum limit.
 The procedure
of taking the diagonal part of the gauge field $A_\mu$ can be implemented
on the lattice by taking the diagonal part of gauge links $U_\mu$; however 
one can also take the diagonal part of the product
of two or more elementary link variables: that leads, due to the 
non-Abelian nature of the theory, to alternative definitions of the 
Abelian projected fields, which differ by $O(a^2)$ terms.
However, such ambiguities disappear as the continuum limit is approached.

In the following we shall take, as usual, the prescription of starting 
from the elementary link variables: after gauge fixing we take the phases 
${\rm diag} (\phi^1_\mu(n),\phi^2_\mu(n), \dots \phi^N_\mu(n))$
of the diagonal part of each gauge link $U_\mu(n)$, as specified in details in 
Appendix~\ref{takediag}, then we construct the Abelian gauge phases 
$\theta^k_\mu(n)$, following Eq.~(\ref{amudef}), as
\begin{equation}
\theta^k_\mu(n) = \sum_{j = 1}^{k} \phi^j_\mu(n)
\label{thetadef}
\end{equation}
and finally the $k$-th 't Hooft tensor (Abelian plaquette) is constructed as
\begin{equation}
\exp \Big[ i \theta^k_{\mu\nu} (n) \Big] = 
\exp \Big[ i (\theta^k_\mu (n) + \theta^k_\nu (n + \hat \mu) 
- \theta^k_\mu (n + \hat \nu) - \theta^k_\nu (n) ) \Big] .
\label{defabplaq} 
\end{equation}

Regarding the location of magnetic monopoles, it is clear that
the recipe of locating points where one of the coefficients 
$c^k$ vanishes, \ie where two consecutive eigenvalues
of $X(n)$ coincide, cannot be implemented on a discrete
space-time. It is however possible to locate them, in the 
unitary gauge, as the points
from which a net magnetic flux comes out. This is implemented
in the standard  De Grand-Toussaint construction~\cite{degrand},
which looks at the net flux coming out of elementary
three-dimensional lattice cubes. In particular, monopole currents of 
a given type are defined as 
\begin{equation}
m^{k}_\mu = {1 \over 2 \pi} \varepsilon_{\mu\nu\rho\sigma} \hat\partial_\nu \overline
\theta^{k}_{\rho\sigma}
\label{degrand1}
\end{equation}
where $\hat{\partial}_{\nu}$ is the lattice derivative and $\overline \theta^{k}_{\rho\sigma}$ 
is the compactified part of the abelian plaquette phase
\begin{equation}
\theta^{k}_{\mu\nu}= \overline \theta^{k}_{\mu\nu} + 2 \pi n^{k}_{\mu\nu}\ ,
\quad n^{k}_{\mu\nu}\in \mathbf{N}\ , \quad \overline \theta^{k}_{\mu\nu}\in [0,2\pi)\ .
\label{degrand2}
\end{equation}

\subsection{Maximal Abelian Gauge and magnetic monopoles in $SU(N)$
gauge theories}
\label{2C}

The location of monopole currents depends on the choice of the Abelian projection.
A standard, popular choice, is to define Abelian projection based on the so-called
Maximal Abelian Gauge (MAG). For $SU(2)$, MAG is defined as the gauge for which the 
following functional
\begin{equation}
F_{\rm MAG} = \sum_{\mu,n} \mbox{tr} \left(U_\mu(n)
  \sigma_3 U^{\dagger}_\mu(n) \, 
\sigma_3\right)
\label{magsu2}
\end{equation}
has a maximum and in Ref.~\cite{mag_su2} this definition was used for the first time in a numerical study. 
It can be shown by explicit calculation that $F_{\rm MAG}$ is 
proportional, apart from a constant term, to the sum of the squared diagonal 
element of the gauge links,
\begin{equation}
F_{\rm MAG} = \sum_{\mu,n} 2 \left( |U_\mu(n)_{11}|^2
+ |U_\mu(n)_{22}|^2 -1 \right)
\end{equation}
and that, at the same time, on stationary points of $F_{\rm MAG}$ the 
traceless and hermitean operator
$\sum_\mu \left[ U_\mu(n) \sigma_3 U^\dagger_\mu(n)
+U^\dagger_\mu(n-\mu) \sigma_3 U_\mu(n-\mu)\right] $
is diagonal and can be related to the local operator 
 $X^{\rm MAG}$ which defines the Abelian projection.
In particular, only on MAG fixed configurations, $X^{\rm MAG}$ 
takes the following explicit expression in terms of 
the gauge links
\begin{equation}
X^{\rm MAG}(n) = \sum_\mu \left[ U_\mu(n) \sigma_3 U^\dagger_\mu(n)
+U^\dagger_\mu(n-\mu) \sigma_3 U_\mu(n-\mu)\right] \, ,
\label{Xdef}
\end{equation}
while in a generic gauge,
connected
to MAG by the local gauge transformation $G(n)$, it is 
defined as
$G(n) 
X^{\rm MAG}(n)
G(n)^\dagger$, which makes it a local
adjoint operator by construction. Notice that this means that
the explicit form of the adjoint field is not known apriori,
but only after MAG has been fixed.

The usual rationale behind the use of the MAG projection is that, since the sum of the 
squared diagonal elements is maximized, abelian projected fields retain most of the 
original Yang-Mills dynamics (Abelian Dominance).
On the other hand, such choice is supported by the empirical fact that the physical properties of MAG monopoles 
show a negligible dependence on the lattice ultraviolet (UV) cutoff.
An important property of the Maximal Abelian Gauge is that magnetic currents, 
defined through the violation of the non-abelian Bianchi identities, are correctly
identified and magnetic charge obeys the correct Dirac quantization 
condition~\cite{nabianchi} (see also Ref.~\cite{bdd} for a related numerical study).
In view of this, in the following we shall adopt the Maximal Abelian Gauge. 
However, the extension of the definition to the generic $SU(N)$ gauge group 
reveals ambiguous and various possibilities exist: in the following we shall
discuss which of them are suitable for a correct definition of magnetic charges, 
in the light of the arguments given in Sections~\ref{2A} and \ref{2B}.

A possible, straightforward generalization for $SU(3)$, introduced in 
Ref.~\cite{mag_su3} and usually adopted
in the literature~\cite{mag3_s,mag3_1,mag3_2,mag3_3},
 is to maximize the sum of the 
squared diagonal elements of all gauge links. The corresponding
functional, in the case of $SU(3)$, reads:
\begin{equation}
\begin{aligned}
F^{SU(3)}_{\rm MAG} &=& \sum_{\mu,n} \left( 
\mbox{tr} \left( U_\mu(n) \lambda_3 U^{\dagger}_\mu(n) \, \lambda_3\right) +
\mbox{tr} \left( U_\mu(n) \lambda_8 U^{\dagger}_\mu(n) \, \lambda_8\right)
\right) \\
&=& 2 \sum_{\mu,n} \left( |(U_\mu(n))_{11}|^2 +  |(U_\mu(n))_{22}|^2 +  
|(U_\mu(n))_{33}|^2 - 1 \right)
\label{magsu3_1}
\end{aligned}
\end{equation}
where $\lambda_3 = {\rm diag}(1,-1,0)$ and 
$\lambda_8 = {\rm diag}(1/\sqrt{3},1/\sqrt{3},-2/\sqrt{3})$. 
This extension seems ideally suited 
for studies regarding Abelian dominance, 
however it has the problem that no natural operator exists,
transforming in the adjoint representation, 
which is diagonal in this gauge, therefore it does not seem 
to be well suited 
for an extension of MAG Abelian projection
to $SU(N)$. We will discuss in more detail the problem later in 
this Section.

An alternative possibility, suggested for $SU(3)$ in 
Refs.~\cite{stack_1,stack_1b,stack_2}
(where it is called ``generalized MAG''), is the following
\begin{equation}
\tilde F_{\rm MAG} = \sum_{\mu,n} 
\mbox{tr} \left( U_\mu(n) \tilde\lambda U^{\dagger}_\mu(n) \, \tilde\lambda \right) \,\, ; \quad \tilde \lambda = 
{\rm diag} (\tla_1,\tla_2, \dots \tla_N) \, ,
\label{magsu3_2}
\end{equation}
where $\tilde \lambda$ is a generic element of the Cartan subalgebra.
For appropriate choices of $\tla$, the maximization of the functional 
in Eq.~(\ref{magsu3_2}) defines the following diagonal operator:
\begin{equation}
\tilde X(n) = \sum_\mu \left[ U_\mu(n) \tilde\lambda U^\dagger_\mu(n)
+U^\dagger_\mu(n-\mu) \tilde\lambda U_\mu(n-\mu)\right] \, .
\label{XdefN}
\end{equation}
Suppose indeed that we have reached an extremum for $\tilde F_{\rm MAG}$, then its variation
for any infinitesimal gauge transformation must vanish. Let us take a particular 
transformation which is non-trivial in a single site,
$G(n) \approx {\rm Id} + i\ \varepsilon\ \Lambda$, then the extremum condition reads:
\begin{equation}
0={\rm tr} \left (G(n) \tla G(n)^\dagger \tilde X(n) \right) - 
{\rm tr} \left( \tla \tilde X(n) \right)
\approx i\ \varepsilon\ {\rm tr} \left( [\Lambda,\tla] X(n) \right)  \, .
\label{extremum}
\end{equation}
A non-trivial condition on $X(n)$ applies if $\Lambda$ is not in the 
Cartan subalgebra, in particular, taking alternatively  
$\Lambda_{ij}^{(kl)} = \delta^k_i \delta^l_j  + \delta^l_i \delta^k_j$
and   
$\Lambda_{ij}^{(kl)} = i (\delta^k_i \delta^l_j  - \delta^l_i \delta^k_j)$,
we obtain {(no summation over repeated indices is intended) }
\begin{equation}
(\tla_k - \tla_l)\ (\tilde X(n)_{kl} \pm \tilde X(n)_{lk}) = 0 \, ,
\end{equation}
which implies $\tilde X(n)_{kl} = 0$ for every $k \neq l$,  unless
$\tla_k = \tla_l$. Therefore $X(n)$ is diagonal if $\tla$ has no pair
of coinciding eigenvalues.
If $\tla$ is expanded over the basis of fundamental weights $\phi_0^k$ 
(see Eq.~(\ref{fwdef}))
\begin{equation}
\tla = \sum_{k = 1}^{N-1} b^k \phi_0^k 
\label{tla1}
\end{equation}
such condition reads
\begin{equation}
\tla_i - \tla_j = \sum_{k = i}^{j-1} b^k \neq 0
\label{tla2}
\end{equation}
for any $i < j$, which implies various constraints, including the fact 
that none of the $b_k$ can vanish. Such constraints can be better specified 
if we now require a given ordering for the eigenvalues of the diagonal operator 
$\tilde X(n)$, for instance the one in Eq.~(\ref{ordering}), 
at least around the perturbative vacuum, where $U_\mu(n) \simeq 1$
and $\tilde X(n) \approx \tla$. 
That implies $\tla_i > \tla_{i+1}$, hence, from Eq.~(\ref{tla2}),
we obtain that $b^k > 0\ $ $\forall\ k$. 

Following the discussion in Sections~\ref{2A} and \ref{2B}, magnetic 
monopoles of type $k$ will be located at points where the coefficient 
$\tilde c^k$ vanishes in the expansion 
$\tilde X (n) = \sum_{k = 1}^{N-1} \tilde c^h(n) \phi_0^h$;
each monopole species will be associated to a particular $U(1)$ residual 
subgroup. For configurations which are close to the perturbative vacuum, 
$U_\mu(n) \simeq 1$, we have $c^k(n) \sim b_k$, hence no monopole 
will appear in any $U(1)$ subgroup, since $b_k \neq 0\ $ $\forall k$; the 
appearance of a monopole therefore requires a non-perturbative fluctuation
of gauge fields, as it is naturally expected.
These considerations lead us to fix a well defined choice for $\tla$, that 
will be adopted in the following: while no particular reason exist to put 
further constraints on $\tla$, apart from $b^k > 0\ $ $\forall\ k$, it is 
clear that a choice for which all coefficients are equal treats all monopole 
species symmetrically, hence it seems preferable.
Therefore we will fix $b^k = 1$ for every $k$, ending up with
the following definition of $\tla$
\begin{equation}
\tla = \sum_{k = 1}^{N-1} \phi_0^k = {\rm diag} 
\left( \frac{N-1}{2}, \frac{N-1}{2} -1, \frac{N-1}{2} -2, \dots, 
-\frac{N-1}{2} \right) \, .
\label{fixtla}
\end{equation}
To better appreciate such choice, consider that, in case one of the 
coefficients is much smaller than the others, than the appearance 
of a monopole-like defect in the corresponding subgroup may be induced 
also by small scale fluctuations, leading to possible ambiguities, 
especially in a lattice setup where one would like to distinguish 
true non-perturbative fluctuations from artifacts at the UV scale.
That also clarifies the advantage of the Abelian projection based 
on MAG, with respect to other possible projections. 
Consider, for instance, the Abelian projection based on the diagonalization 
of the plaquette operator, taken, e.g., in the $12$ plane:
\begin{equation}
X_{12}(n) = -i \left(\Pi_{12} - \Pi_{12}^\dagger \right) + 
 i\frac{1}{N} {\rm tr} \left(\Pi_{12} - \Pi_{12}^\dagger \right)\, .
\end{equation}
In the limit of small fields, $U_\mu(n) \simeq 1$,  we have 
$X_{12} (n)= \sum_{k = 1}^{N-1} c_{12}^k(n) \phi_0^k \simeq 0$, 
meaning that $c_{12}^k \simeq 0$ for every $k$. Hence we expect that 
the Abelian projection based on the plaquette operator will detect 
many magnetic monopoles even in the limit of small fields: in this 
case the detection procedure will be strongly affected by lattice artifacts,
\ie by UV noise, leading, e.g.,  to a wrong scaling of the monopole density 
in the continuum limit. The same argument applies to any other Abelian 
projection based on a local adjoint operator vanishing in the limit of 
small gauge fields.
\\

Finally, let us summarize and further clarify the reasons leading us
to the proposed extension of MAG projection for $SU(N)$ gauge theories 
and to discard other possibilities considered in the literature, like 
that based on the functional in Eq.~(\ref{magsu3_1}).
The original aim of Abelian projection is to fix the non-abelian gauge 
symmetry apart from a residual maximal $U(1)^{N-1}$ Abelian subgroup, 
to which $N-1$ electric and magnetic charges can be associated. The need 
for fixing the $(N-1)$ $U(1)$ subgroups simultaneously, \ie by the same
gauge condition, stems from the requirement that each electric or magnetic 
charge be neutral with respect to all other $U(1)$ subgroups.

>From this point, it may seem that fixing the gauge by looking for a maximum 
of the functional in Eq.~(\ref{magsu3_1}) may work equally well, even if no 
diagonal operator is naturally associated to such a choice. Indeed, the 
functional is invariant under gauge transformations belonging to the maximal 
$U(1)^{N-1}$ subgroup, which is therefore well defined on each stationary 
point of the functional. However, the crucial point is that $U(1)^{(N-1)}$ 
must be the only residual symmetry, \ie one should take care of fixing any 
possible additional symmetry, like that under index permutations.

The functional in Eq.~(\ref{magsu3_2}), on which our choice of Abelian projection 
is based, is in general not invariant under local permutations of the color indexes, 
$i \to P(n,i)$, which transform gauge links as follows: 
$U_\mu(n)_{ij} \to U_\mu(n)_{P(n,i)\, P(n + \hat \mu, j)}$.
Indeed, it is easily verified that such transformation is equivalent to 
performing a local modification of $\tla$, corresponding to a permutation of its 
diagonal elements, $\tla_i \to \tla_{P^{-1}(n,i)}$, which changes the value 
of the functional.

The operator in Eq.~(\ref{magsu3_1}) is not invariant under local permutations as 
well, since in general they change the identification of the diagonal elements. 
However, global permutations, \ie with $P(n,i)$ independent of $n$, are a 
residual symmetry of such operator. Such residual symmetry, unfortunately, 
makes it ambiguous to identify in which subgroup a given magnetic monopole appears.
A direct consequence, that will be verified numerically in the next subsection, 
is that when one constructs magnetic currents according to the abelian projected 
phases in such a gauge, magnetic monopoles appear simultaneously on the same site 
and in different $U(1)$ subgroups, meaning that the magnetic charge operators are 
not well identified, \ie they do not commute with each other.

Actually, the particular choice adopted for $\tla$ in Eq.~(\ref{fixtla}), has a 
global symmetry as well, corresponding to the following global permutation: 
$i \to (N - i + 1)$, \ie to the inversion in the location 
of all eigenvalues of $\tla$. 
However, such residual symmetry is harmless, since it sends pairs of adjacent eigenvalues
into pairs of adjacent eigenvalues, hence it changes the location of the $U(1)$ 
subgroups, but leaves them still well identified.

\subsection{Implementation for $SU(3)$ and the detection of independent monopole species.}

In the following we will apply the above considerations to the study of 
magnetic monopoles in the $SU(3)$ pure gauge theory. To summarize, our procedure
for Abelian projection will be to fix the gauge by maximizing the functional
\begin{equation}
\tilde F^{\rm SU(3)}_{\rm MAG} = \sum_{\mu,n} 
\mbox{tr} \left( U_\mu(n) \tilde\lambda U^{\dagger}_\mu(n) \, \tilde\lambda \right) \ ; 
\quad \tla = {\rm diag} (1,0,-1) \, .
\label{magsu3_3}
\end{equation}
Then we take the diagonal part of gauge links,
$U^D_\mu (n) = {\rm diag} (e^{i \phi^1_\mu(n)},e^{i \phi^2_\mu(n)},e^{i \phi^3_\mu(n)})$,
and determine the Abelian phases $\theta^1_\mu(n),\theta^2_\mu(n)$,
corresponding to the two residual $U(1)$ subgroups, following Eq.~(\ref{thetadef}), 
which for $SU(3)$ reads $\theta^1_\mu(n) = \phi^1_\mu(n)$ and 
$\theta^2_\mu(n) = - \phi^3_\mu(n)$ (remember that $\sum_i \phi^i_\mu(n) = 0$). 
The numerical algorithm adopted to maximize the functional in Eq.~(\ref{magsu3_3}) 
and that used to extract $U^D_\mu(n)$ are illustrated respectively in 
Appendix~\ref{magalg} and \ref{takediag}.

Starting from $\theta^1_\mu(n)$ and $\theta^2_\mu(n)$, we determine the two monopole 
currents $m^{1}_\mu$ and $m^{2}_\mu$, following the De Grand-Toussaint 
method~\cite{degrand} (see Eqs.~(\ref{degrand1}) and (\ref{degrand2})). Monopole 
currents form closed loops, since $\hat\partial_\mu m^k_\mu=0$, and we will be 
interested in particular in monopole loops with a non-trivial wrapping around 
the Euclidean temporal direction, which can be identified with thermal 
monopoles~\cite{chezak,bornya92,ejiri,monden}, and whose properties 
in the deconfined phase of $SU(3)$ gauge theory will be studied in Section~\ref{monden}.

It is interesting to write down a continuum expression for 
the functional in Eq.~(\ref{magsu3_3}).
A straightforward computation yields, apart from a constant term,
\begin{equation}
\tilde F^{\rm SU(3)}_{\rm MAG} = 
-\frac{1}{2}\sum_{\mu, n}\Big\{|U_{\mu}(n)^{12}|^2+|U_{\mu}(n)^{12}|^2+
|U_{\mu}(n)^{23}|^2+|U_{\mu}(n)^{32}|^2
+4\Big[|U_{\mu}(n)^{13}|^2+|U_{\mu}(n)^{31}|^2\Big]\Big\}
\end{equation}
which in the continuum limit, 
$U_{\mu}(n)^{j\ell}\approx \delta^{j\ell}+iag A_{\mu}^{j\ell}$,
becomes
\begin{equation}
\tilde F^{\rm SU(3)}_{\rm MAG} \approx
-\frac{1}{2}\sum_{\mu}\int \mathrm{d}^4 x\Big\{|A_{\mu}(x)^{12}|^2+|A_{\mu}(x)^{12}|^2+
|A_{\mu}(x)^{23}|^2+|A_{\mu}(x)^{32}|^2
+4\Big[|A_{\mu}(x)^{13}|^2+|A_{\mu}(x)^{31}|^2\Big]\Big\} \, .
\end{equation}
If we adopt the usual notation
$A_{\mu}=(\lambda_{a}/2) A_{\mu}^a$, where $\lambda_a$ are the 
Gell-Mann matrices, we obtain 
\begin{equation}
\tilde F^{\rm SU(3)}_{\rm MAG} \approx
-\frac{1}{2}\sum_{n}\int \mathrm{d}^4 x\Big\{(A_{\mu}(x)^1)^2+(A_{\mu}(x)^2)^2+
(A_{\mu}(x)^6)^2+(A_{\mu}(x)^7)^2
+4\Big[(A_{\mu}(x)^4)^2+(A_{\mu}(x)^5)^2\Big]\Big\} \; ;
\end{equation}
the Higgs field defining the Abelian projection is 
oriented like $\tla$ in the gauge fixed configuration 
(see also Ref.~\cite{stack_1} for a derivation in the case of a
more general choice for $\tla$).
For comparison, we report also the continuum expression corresponding to
the standard extension 
of MAG to $SU(3)$, Eq.~(\ref{magsu3_1}),
\begin{equation}
\tilde F^{\rm SU(3)}_{\rm MAG} \approx
-\frac{1}{2}\sum_{n}\int \mathrm{d}^4 x\Big\{(A_{\mu}(x)^1)^2+(A_{\mu}(x)^2)^2+
(A_{\mu}(x)^4)^2+(A_{\mu}(x)^5)^2
+ (A_{\mu}(x)^6)^2+(A_{\mu}(x)^7)^2 \Big\} \, , 
\end{equation}
whose maximization corresponds to the minimization of the non-diagonal part
of the gauge field, but has no Higgs field associated with it.

To conclude 
the present Section, and before starting a detailed study of thermal 
monopole properties, we would like to discuss numerical data showing that our choice 
of Abelian projection indeed leads to independent monopole currents. In order to give 
a quantitative measure of such ``independence'', let us call $\rho_A$ ($\rho_B$) the 
probability that a given three-dimensional cube of the lattice is pierced by monopole 
current $m_\mu^A$ ($m_\mu^B$), and by $\rho_{AB}$ the probability that the cube is 
pierced both by current $m_\mu^A$ and current $m_\mu^B$. If the two monopole currents
were completely independent of each other, than one would expect 
\begin{equation}
\frac{\rho_{AB}}{\rho_A \rho_B} = 1 \, ,
\label{ratioover}
\end{equation}
\ie the probability of two coinciding currents should be equal to the product of the 
respective probabilities. Actually, even if the two magnetic charges are independent, 
each of them being neutral with respect to the $U(1)$ group of the other charge, we 
still expect some correlations of physical origin, which is due to interactions 
induced by the non-Abelian degrees of freedom: such interactions will be studied in 
more detail in Section~\ref{monden}, where we analyze spatial correlations among 
thermal monopoles. In any case, even considering such interactions, the probability 
of having two different monopoles exactly on the same lattice cube should not be 
particularly enhanced, especially when, approaching the continuum limit, the volume 
of the single cube shrinks to zero. Therefore the ratio in Eq.~(\ref{ratioover}) 
should stay of $O(1)$.

\begin{figure}[h!]
\begin{center}
\includegraphics*[width=0.65\textwidth]{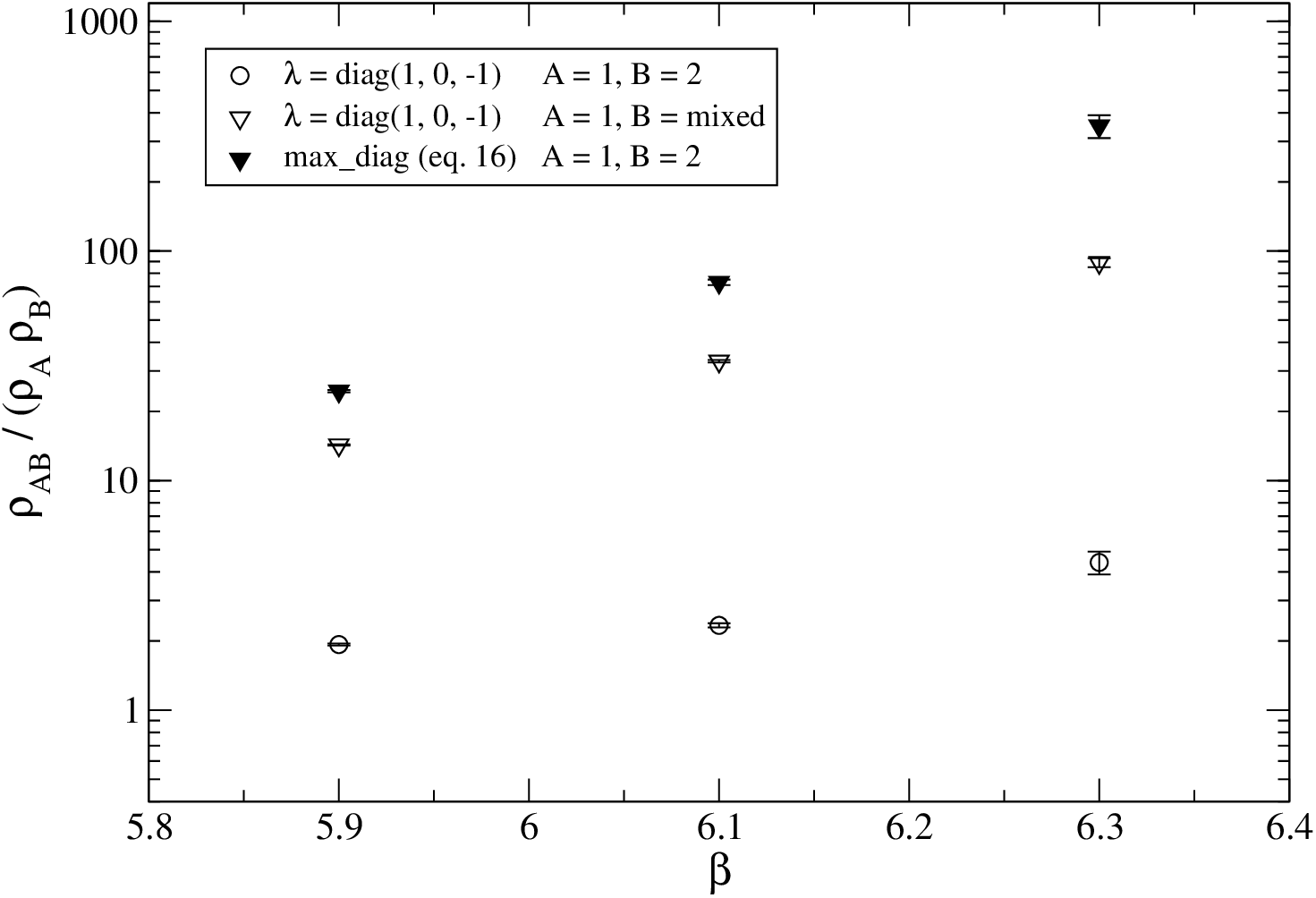}
\vspace{-0.cm}
\caption{Probability of two coinciding monopole currents, normalized
by the corresponding single current probabilities (see Eq.~(\ref{ratioover})),
for different values of $\beta$ and 
for different choices of the monopoles currents, corresponding 
to the Abelian projections specified by Eq.~(\ref{magsu3_3}) or by 
Eq.~(\ref{magsu3_1}) (${\rm max\_diag}$). $A = 1$ and $B = 2$ correspond
to $\theta^1_\mu(n) = \phi^1_\mu(n)$
and $\theta^2_\mu(n) = - \phi^3_\mu(n)$ respectively, while 
$B = {\rm mixed}$ corresponds to $\theta^2_\mu(n) = \phi^2_\mu(n)$ (see text).
}
\label{overlap} 
\vspace{-0.cm}
\end{center}
\end{figure}

We have studied the ratio reported in Eq.~(\ref{ratioover}) for the $SU(3)$ pure 
gauge theory, with the Wilson (plaquette) action discretization, on a $16^4$ lattice, 
for a few values of the inverse gauge coupling $\beta$ and for different choices of 
the monopole currents, $m_\mu^A$ and $m_\mu^B$. Results are reported in Fig.~\ref{overlap}.
The first choice corresponds to the two independent monopole currents identified by 
the definition of MAG adopted in the rest of this study, see Eq.~(\ref{magsu3_3}):
the ratio is $O(1)$ for all explored values of $\beta$, as expected for independent
currents. 

It is interesting to notice that, even in this gauge, if the second current is 
constructed making use of the wrong abelian phase, for instance taking 
$\theta^2_\mu(n) = \phi^2_\mu(n)$, such property is completely lost and the ratio
in Eq.~(\ref{ratioover}) is of the order of $10^1$ - $10^2$: in the gauge specified 
by Eq.~(\ref{magsu3_3}) such phase receives contribution from both $U(1)$ subgroups, 
hence it defines a magnetic current which strongly overlaps with both correct currents. 
As a matter of fact, if one of the two correct monopole currents pierces a given 
lattice cube, there is a probability larger than 50\% that a monopole current, 
corresponding to the fake abelian phase, is found on the same cube. 

Last choice corresponds to the gauge fixing specified by the functional in 
Eq.~(\ref{magsu3_1}) and to the currents associated again to 
$\theta^1_\mu(n) = \phi^1_\mu(n)$ and $\theta^2_\mu(n) = - \phi^3_\mu(n)$. 
The overlap between the two currents is clearly visible, with 
$\rho_{AB}/(\rho_A \rho_B)$ of the order of $10^1$ - $10^2$ over the explored 
range of $\beta$'s. 
The probability that a monopole current of a given type is found in a
cube where another current type has already been found is of the order of 20-30\% over 
the whole range of explored $\beta$ values. In this case, similar results are 
obtained considering any pair of abelian phases, since all diagonal elements 
are treated symmetrically.

A final comment regards the problem of gauge fixing ambiguities. It is well known
that functionals like those in Eqs.~(\ref{magsu3_1}) and (\ref{magsu3_2}) possess 
many local maxima, corresponding to different gauge fixed configurations, \ie 
different Gribov copies. A usual choice, in order to fix the gauge unambiguously, 
is to define it as that corresponding to the global maximum of the functional, 
even if finding it may result computationally expensive. Such strategy is well 
justified for the functional in Eq.~(\ref{magsu3_1}) and for questions related 
to Abelian dominance, \ie regarding the possibility of reproducing physical 
properties of the non-Abelian theory by the diagonal part of gauge fields only.

In the case of our interest, however, every local maximum of the functional 
$\tilde F_{\rm MAG}$ in Eq.~(\ref{magsu3_2}) will lead to a well defined 
diagonal operator $\tilde X(n)$, hence to a legitimate Abelian projection, on 
the same footing with other Gribov copies, including the global maximum of the 
same functional.
Different Gribov copies will 
lead to different adjoint operators $\tilde X(n)$, hence 
to different Abelian projections and 
different monopoles.
In that sense, numerical studies which look for the global maximum of the MAG 
functionals can be considered as an important tool to reveal the systematic 
uncertainties linked to such choice.

Taking the Gribov copy which is found first by the maximization procedure, 
starting from the original configuration sampled by the Monte-Carlo algorithm, 
can be considered as a practical criterion which leads to a correct behavior of 
the monopole properties in the continuum limit~\cite{monden}: we will follow such 
prescription in our study of thermal monopoles for $SU(3)$. Accurate studies have 
been performed for $SU(2)$ thermal monopoles, adopting simulated annealing 
procedures in order to get as close as possible to the global maximum of the 
MAG functional~\cite{bornyakov, born1, born2}: apart from a 20\% difference in the overall
density of thermal monopoles, no other significant differences have been 
revealed, regarding the main physical properties of thermal monopoles.

It would be interesting, in future studies, to consider the implementation 
for $SU(N)$ gauge theories of different gauge conditions which, while sharing
with the MAG the property of being safe from ultraviolet fluctuations, are 
also free of Gribov copy ambiguities. One possibility
could be the so-called Laplacian gauge, see Refs.~\cite{lapla1,lapla2}.

\section{Numerical simulations and results}
\label{monden}

Monopole currents form closed loops, which may wrap by periodicity around the 
lattice torus. While in the confined phase wrappings in both spatial and temporal 
directions take place, usually associated to the presence of a large, percolating 
cluster of currents, in the deconfined phase the non-trivial wrappings survive 
only in the Euclidean, periodic time direction, which is associated to the thermal 
properties of the theory.

Such trajectories with non-trivial temporal wrappings can be associated with thermal 
objects populating the finite $T$ medium~\cite{bornya92,ejiri,chezak}, in analogy 
with the path-integral representation of the partition function of a system of 
quantum particles. For that reason they have been directly related to the magnetic
component of the deconfined plasma~\cite{chezak}, constituted by thermal abelian 
monopoles evaporating from the low $T$ magnetic condensate. Many properties of thermal
monopoles have been studied in subsequent 
works~\cite{monden,moncon,dyons,born1,born2,born3,braguta}, 
which are of interest for 
the comprehension of the deconfined state of Yang-Mills theories and of the 
confinement/deconfinement mechanism itself. Such studies will be extended to 
$SU(3)$ in the present Section.

We summarize the main steps for identifying thermal monopoles on a given gauge 
configuration. We first look 
for monopole currents piercing a given time slice of the lattice (\eg, at $t = 0$)
in the temporal direction, then we follow the current around the lattice and keep 
trace of the number of temporal wrappings the current makes before going back to 
the original detection point. Currents wrapping in the positive (negative) 
direction are associated to monopoles (antimonopoles). If the current
wraps only one time, then the spatial location of the thermal monopole
on the starting time slice is well defined, apart from possible ambiguities 
at the UV scale, due to short range fluctuations of the trajectory.
If the current wraps 2 or more times, then it can be associated 
to 2 or more thermal monopoles undergoing a cyclic permutation as 
they go around the thermal cycle: such trajectories are typical of the
path integral representation of a system of identical, 
bosonic particles.  Possible ambiguities in the identification procedure
described above can occurr when two trajectories cross at the same point:
it is not clear if they represent a single trajectory with a double wrap
or not; however, from a practical point of view, 
the presence of two currents in the same lattice cube
is very rare event, hence of no statistical significance.

In the following, we will show and discuss numerical results obtained
by simulating $SU(3)$ with the Wilson plaquette action (making
use of a standard combination of heat-bath \cite{cabibbo}
and over-relaxation \cite{creutz} updates). The temperature 
$T = 1/(L_t a(\beta))$ has been tuned both by changing the 
number of temporal lattice sites, $L_t$, at fixed UV scale 
(with $L_t$ ranging from 4 to 11) and by tuning the inverse bare 
coupling $\beta$ at fixed $L_t$. A number of different spatial sizes 
have been explored, ranging from $24^3 \times L_t$ to $48^3 \times L_t$. 
In order to obtain the physical value of the temperature, in units 
of $T_c$, we have exploited the non-perturbative determination
of $a(\beta)$ and the critical $\beta$ values for various values
of $L_t$ reported in Ref.~\cite{karsch_thermo}. 
For each parameter set we have measured thermal monopole properties
on a number of decorrelated gauge configurations ranging from 
a few hundreds up to a few thousands.

A first quantity that we will look at is the total density of 
thermal monopoles as a function of $T$. Results for $SU(2)$ 
show that the density increases with $T$, as expected for 
a particle-antiparticle gas, but with a typical logarithmic
suppression which can be related to the temperature behavior
of the magnetic coupling and shows that monopoles degrees of 
freedom become irrelevant, with respect to gluons, 
in the high $T$, perturbative regime, while their role 
is more and more significant as the deconfining temperature $T_c$
is approached from above. 

Further information will be obtained by looking at the density 
of trajectories with multiple wrappings. As discussed above, 
such trajectories can be interpreted in terms of the exchange of identical
particles: they are strongly suppressed at high $T$, where the 
system is quasi-classical, and become statistically relevant at 
low $T$. The analysis of Ref.~\cite{moncon}, which is based on the 
analogy with a free boson gas, has shown that for $SU(2)$ 
the distribution of multiple wrapping trajectories as a 
function of $T$ can be used to detect the point where thermal monopoles
seem to condense, and that such point coincides within errors
with $T_c$, giving support to a confinement mechanism based on the 
condensation of magnetic charge (dual superconductor mechanism). 
The same analysis will be repeated for $SU(3)$ in Section~\ref{sub:moncon}.

Finally, in Section~\ref{sub:monint}, we will analyize the spatial
correlations among thermal monopoles. Already in $SU(2)$, where a single
monopole species exists, such correlations furnish interesting
information, showing the presence of Coulomb-like, screened interactions
among monopoles and antimonopoles, with an effective magnetic coupling 
which grows with $T$. For $SU(3)$, apart from verifying the presence 
of analogous interactions among monopoles and antimonopoles of the 
same kind, we will have the possibility of investigating 
the correlations between the two different species, which
are related to the non-Abelian nature of the theory and will reveal
to be highly non-trivial.

\subsection{Monopole density}
\label{sub:monden}

\begin{figure}[t!]
\begin{center}
\includegraphics*[width=0.65\textwidth]{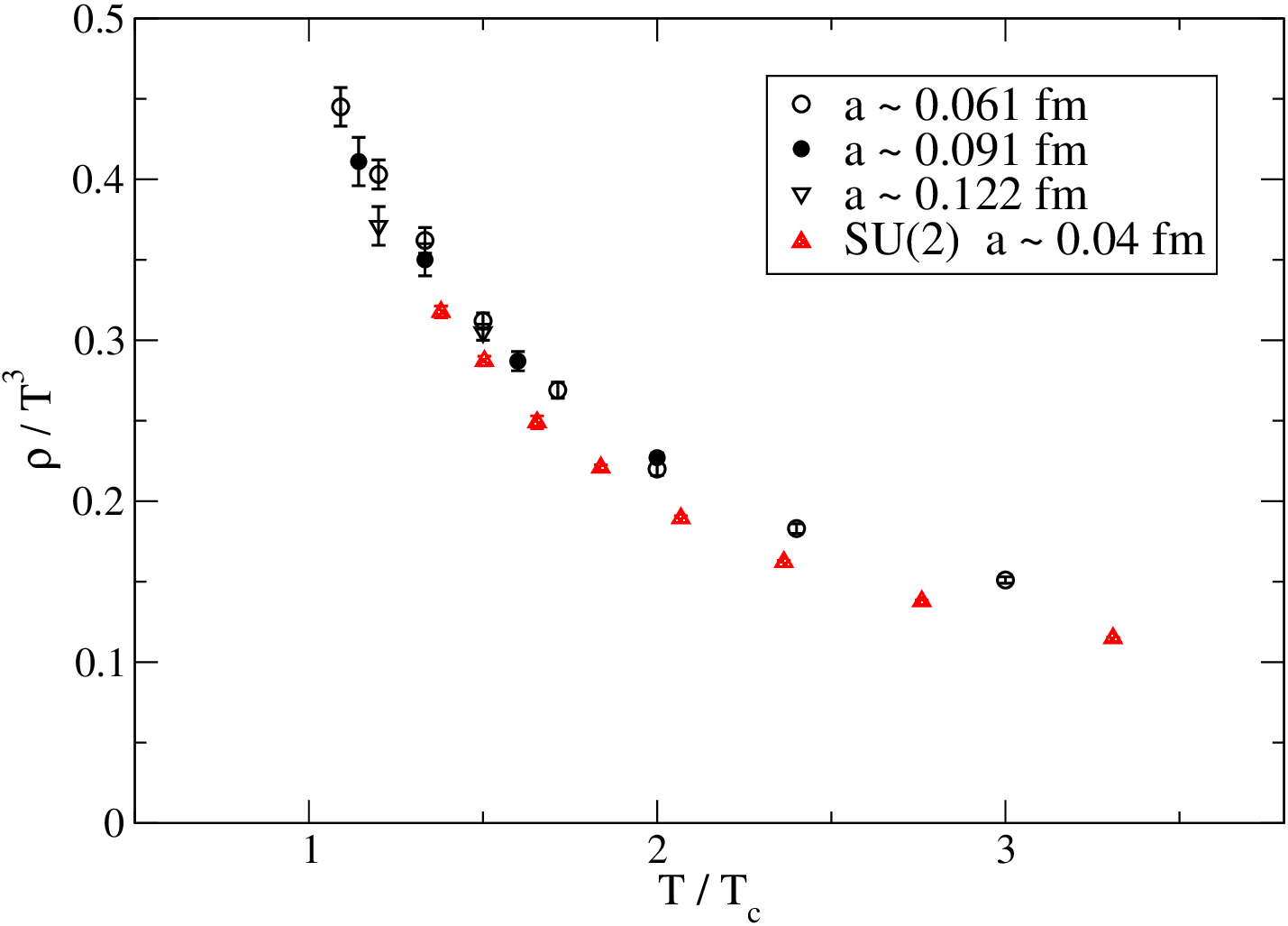}
\vspace{-0.cm}
\caption{Total density, normalized by $T^3$, of thermal monopoles of the first
species, determined for different values of $T$ and of the lattice spacing $a$.
We show for comparison also data obtained for $SU(2)$ from 
Ref.~\cite{monden}.
}
\label{densl3_phys} 
\vspace{-0.cm}
\end{center}
\end{figure}

We define the total density of thermal monopoles of a given species
as~\cite{chezak,bornya92,ejiri}
\begin{equation}
\rho = \frac{\left< \sum_{\vec{n}} \left| N_{wrap}(m_0(\vec{n},t))
\right| \right> }{V_s}
\label{densdef}
\end{equation}
where $N_{wrap}(m_0(\vec{n},t))$ is the winding number in the 
temporal direction of the monopole current $m_0$ 
initially detected at the lattice site $(\vec{n},t)$, the sum is 
over all spatial sites at a given time slice $t$ and 
$V_s = (L_s a)^3$ is the spatial volume. 

It is convenient to define the following dimensionless ratio
\begin{eqnarray}
\frac{\rho}{T^3} = 
\frac{\left< L_t^3 \sum_{\vec{n}} \left| N_{wrap}(m_0(\vec{n},t))
\right| \right> }{L_s^3} \, ,
\end{eqnarray}
which for a gas of free quantum particles and antiparticles of mass
$m$ should tend to a constant as $T \gg m$. 

In Fig.~\ref{densl3_phys} we show $\rho/T^3$ for the first 
monopole species and for different temperatures and lattice
spacings, results obtained for the second species are compatible
within errors, as expected from our choice of $\tla$ in 
Eq.~(\ref{magsu3_3}). For comparison, we also report data obtained
for $SU(2)$ in Ref.~\cite{monden}.
We notice a good scaling to the continuum limit. Data for $SU(2)$
seem to stay slightly below the $SU(3)$ ones,
however it is important
to stress that the comparison is made at fixed $T/T_c$ and that
$T_c$ is about 10\% higher for $SU(2)$ than for $SU(3)$: if data
were compared at fixed $T/\sqrt{\sigma}$, where $\sigma$ is the string
tension, they would be practically coincident. Therefore, 
$\rho/T^3$ seems to reach slightly 
higher values, around $T_c$, for $SU(3)$ than for $SU(2)$, but
just because $T_c$ is lower for $SU(3)$ and $\rho/T^3$ is a 
decreasing function of $T$.

\begin{figure}[t!]
\begin{center}
\includegraphics*[width=0.65\textwidth]{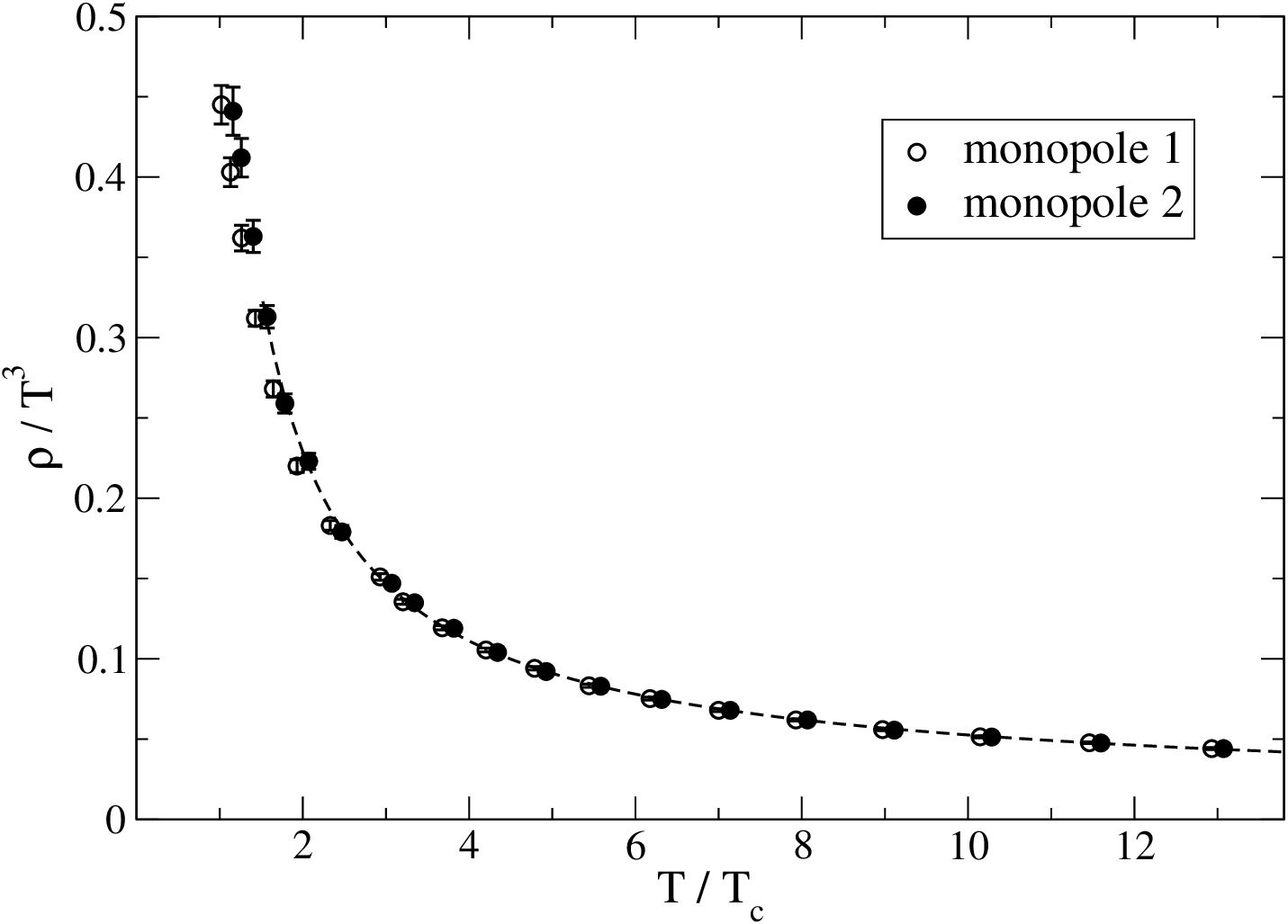}
\vspace{-0.cm}
\caption{$\rho(T)/T^3$ as a function of $T/T_c$ and for the
two different monopole species. Data, which have been slightly split
on the $T/T_c$ axis for the sake of readability, have been
  obtained on a $48^3 \times L_t$ lattice, with variable $L_t$ and
  at $\beta = 6.3368$ (first 9 points), and with variable $\beta$ 
and $L_t = 4$ (last 11 points).
The dashed line is a best fit to Eq.~(\ref{fitlog}).}
\label{dens_hight} 
\vspace{-0.cm}
\end{center}
\end{figure}

The dependence of $\rho/T^3$ on $T$ is best appreciated 
from Fig.~\ref{dens_hight}, where the densities of both monopole
species are shown over an extended temperature range.
It is clear that $\rho/T^3$ does not approach a constant behavior,
so that, like for $SU(2)$~\cite{monden}, a 
description of thermal monopoles as a gas of free particles 
is not appropriate, even at asymptotically high $T$, in agreement
with a scenario based on the electric-magnetic duality~\cite{shuryak},
according to which the high $T$ phase 
of Yang-Mills theories is electrically dominated, while the 
magnetic component is strongly interacting.

Explicit predictions can be done for $\rho/T^3$, based on perturbative and 
dimensional reduction considerations, leading to a behaviour proportional to 
$g^6$~\cite{giovannangeli,shuryak}, where $g(T)$ is the renormalized 
running coupling, hence a reduction factor for $\rho/T^3$, with respect 
to the free massless particle case, of the order of $1/(\log (T/\Lambda_{eff}))^3$, 
where $\Lambda_{eff}$ is some effective scale. Based on such prediction, we have 
tried to fit data\footnote{Reported fit values have been obtained for the first monopoles species,
but results for the second species are completely equivalent.} 
in Fig.~\ref{dens_hight} according to 
\begin{equation}
\frac{\rho}{T^3} = \frac{A}{(\log (T/\Lambda_{eff}))^\alpha}\, .
\label{fitlog}
\end{equation}
If we consider only $T/T_c \geq 2$ and fix $\alpha = 3$, we obtain
$A = 3.66(7)$, $\Lambda_{eff}/T_c = 0.163(4)$ and 
$\chi^2/{\rm d.o.f.} = 9.5/13$. If we instead leave $\alpha$ as 
a free parameter, we get $\alpha = 3.01(33)$ and
$\chi^2/{\rm d.o.f.} = 9.6/12$, in very good agreement 
with the perturbative prediction.

\begin{figure}[h!]
\begin{center}
\includegraphics*[width=0.65\textwidth]{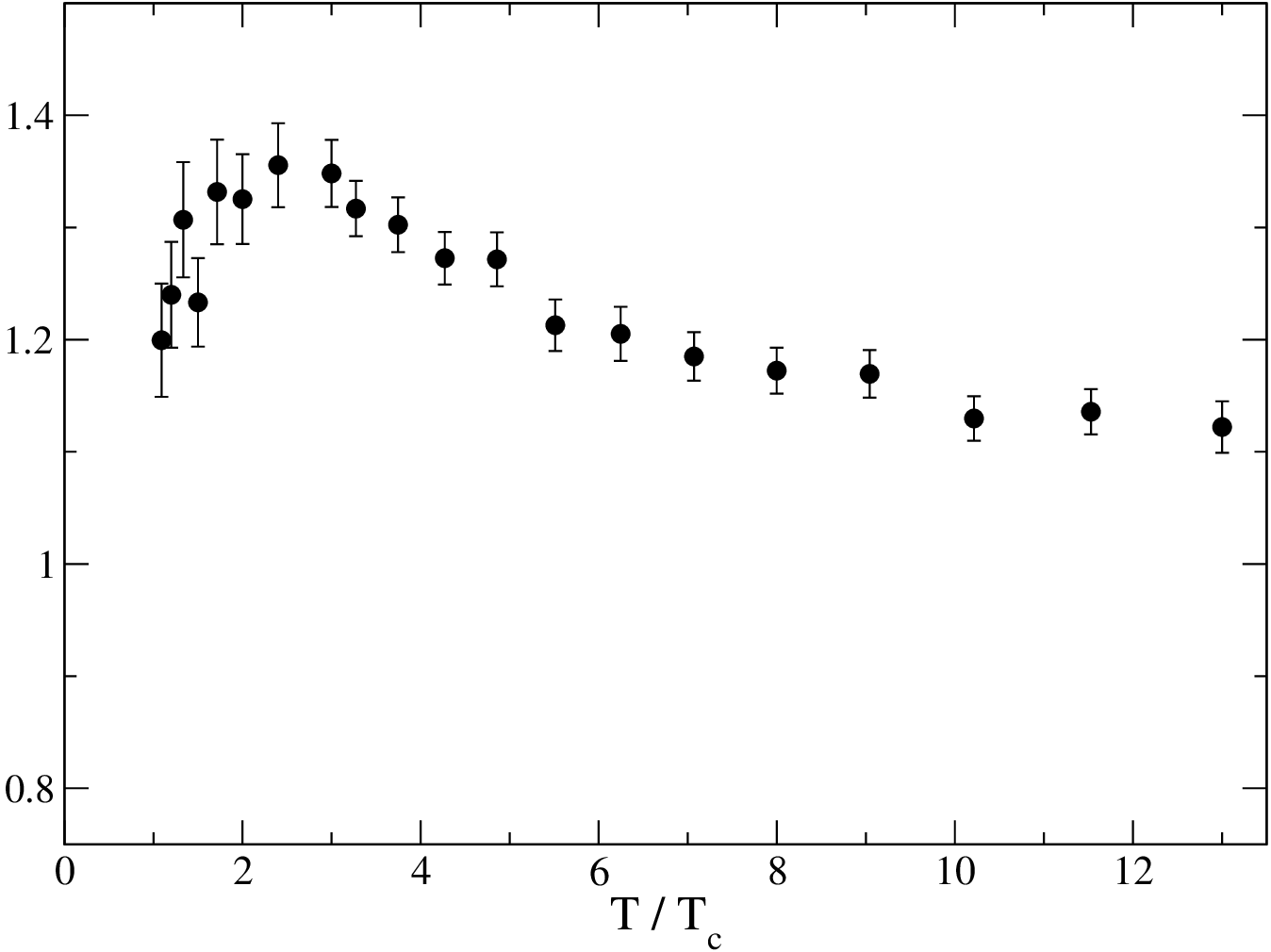}
\vspace{-0.cm}
\caption{Ratio $\rho_1/\rho_2$ of the monopole densities obtained by using different gauge fixing conditions:
$\rho_1$ is determined by using the functional in Eq.~\eqref{magsu3_3}, 
while $\rho_2$ is determined by using the functional in Eq.~\eqref{magsu3_1}.
} 
\label{l3suall} 
\vspace{-0.cm}
\end{center}
\end{figure}

Finally, we would like to make a direct comparison
with the results obtained by adopting the standard extension of 
MAG to $SU(3)$ 
based on the functional in Eq.~(\ref{magsu3_1}), like in 
Ref.~\cite{bornsu3}:
that does not permit a proper distinction between the two 
monopole species, it is however interesting to compare the overall
densities.
In Fig.~\ref{l3suall} we report
the ratio of thermal monopole densities, as a function of $T/T_c$, obtained
on our sample of thermalized configurations
in the two cases: the gauge fixing procedure is similar,
in particular for both functionals 
we stop on the local maximum which is first found when starting 
from the Monte-Carlo thermalized configuration.
Results obtained adopting the functional in 
Eq.~(\ref{magsu3_1}) are consistently lower, by 10-30\%, over the whole
range of explored temperatures: this is consistent with the 
results presented in Ref.~\cite{bornsu3}, 
which are based on Eq.~(\ref{magsu3_1}) and provide evidence
for a thermal monopole density in $SU(3)$ lower than 
that obtained for $SU(2)$, while our results show instead 
that they are 
practically equal.

\subsection{Monopole condensation}
\label{sub:moncon}

Trajectories with multiple temporal wrappings can be related to the 
nature and properties of monopoles as identical quantum particles,
as follows from the interpretation of the set of monopole trajectories,
extracted from a given gauge field, as a possible configuration 
of the Euclidean path integral of an ensemble of identical particles.
Indeed, the path integral of $N$ identical particles at thermal equilibrium is made
up of path configurations which are periodic apart from a possible permutation 
of the $N$ particles, meaning that each configuration presents in general $M$ 
closed paths, with $M \leq N$ and the $j$-th path wrapping $k_j$ times, in 
such a way that $\sum_{j = 1}^M k_j = N$: such configuration 
corresponds to a permutation made up of $M$ cycles of sizes $k_1,k_2,\ \dots\ k_M $. 

When effects related to quantum statistics are negligible, \ie when 
the system is close to the Boltzmann approximation (like it happens, 
for example, at high $T$), configurations deviating from the identical 
permutation have a negligible weight in the path integral, so that
trajectories presenting multiple wrappings are very rare.
Their statistical weight is instead expected to increase as
quantum effects become more important, and in a well defined, critical
way as one approaches typical phenomena like Bose-Einstein Condensation 
(BEC). A treatment of BEC-like phenomena by a path-integral approach goes
back to the seminal papers by Feynman~\cite{fey1,fey2}, where a path integral 
formulation was applied to describe the superfluid transition in $^4$He.

\begin{figure}[h!]
\begin{center}
\includegraphics*[width=0.65\textwidth]{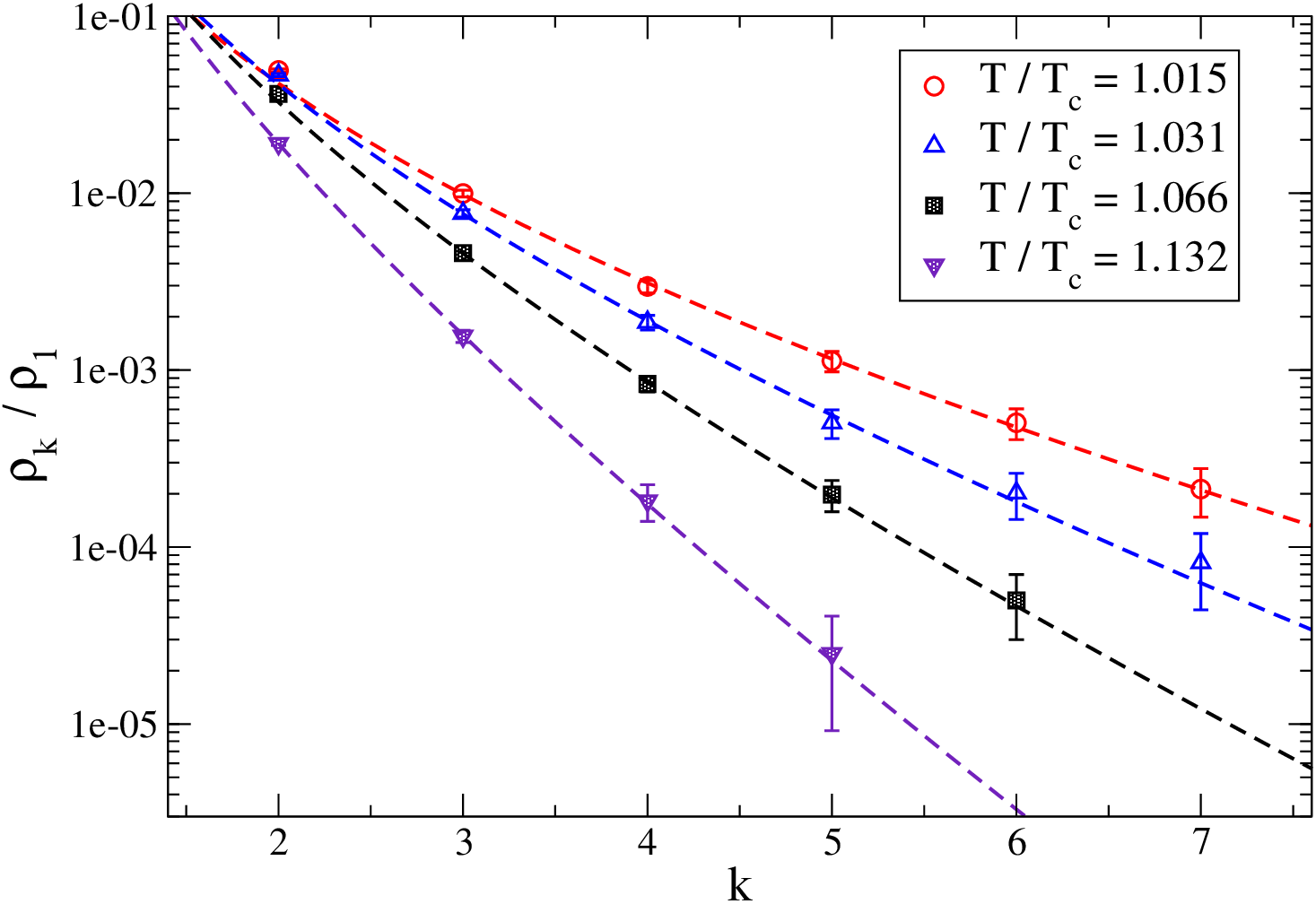}
\vspace{-0.cm}
\caption{Relative density of trajectories with $k$
wrappings as a function of $k$ and for different values 
of $T$, obtained on  a $48^3 \times 8$ lattice at
$\beta = 6.07,\, 6.08,\, 6.10,\,$ and  $6.14$.
Dashed lines correspond to best fits to Eq.~(\ref{rhok}).
}
\label{rhokfig} 
\vspace{-0.cm}
\end{center}
\end{figure}

In particular, for a set a free bosons of mass $m$, one finds that the 
density of paths wrapping $k$ times ($k$-cycles) is given 
by (see, \eg, Refs.~\cite{elser,moncon})
\begin{eqnarray}
\rho_k \equiv \frac{\langle n_k \rangle}{V_s} = \frac{e^{ - \hat\mu k}}{\lambda^3\ k^{5/2}}
\label{rhok}
\end{eqnarray}
where $n_k$ is the number of $k$-cycle in one configuration,
$\lambda=\sqrt{2\pi/(m T)}$ is the De Broglie thermal wavelength,
and the dimensionless quantity 
$\hat \mu$ is related to the usual chemical potential $\mu$ for 
free bosons by $\hat\mu \equiv - \mu/T$, with the constraint 
$\hat\mu \geq 0$ (\ie $\mu \leq 0$). As $\hat\mu \to 0$, \ie as 
the system approaches BEC, higher $k$-cycles become more and more 
frequent and the exponentially suppressed behavior
of $\rho_k$ turns into a critical power law behavior.

In Ref.~\cite{moncon}, 
the distribution of monopole trajectories wrapping
$k$ times has been used, in combination with a simple
ansatz like that in Eq.~(\ref{rhok}), to extract
$\hat \mu (T)$ for the thermal monopoles in the $SU(2)$ gauge theory
and infer that they undergo condensation at a temperature 
$T_{\rm BEC}$ which coincides, within errors, with the 
deconfinement temperature $T_c$. We would like to repeat 
a similar analysis for $SU(3)$.

In Fig.~\ref{rhokfig} we report $\rho_k$, the density of trajectories
wrapping $k$ times, normalized by $\rho_1$, as a function of $k$
for a few values of $T$. Results have been obtained on 
a $48^3 \times 8$ lattice. It is evident that, for each $k$, 
the relative weight of $\rho_k$ rapidly increases as 
$T_c$ is approached from above. Moreover, for each $T$ data can be nicely
fitted according to the simple ansatz in Eq.~(\ref{rhok}), as it 
happens in the $SU(2)$ case, giving us access to the effective
chemical potential $\hat \mu(T)$. 

\begin{figure}[h!]
\begin{center}
\includegraphics*[width=0.65\textwidth]{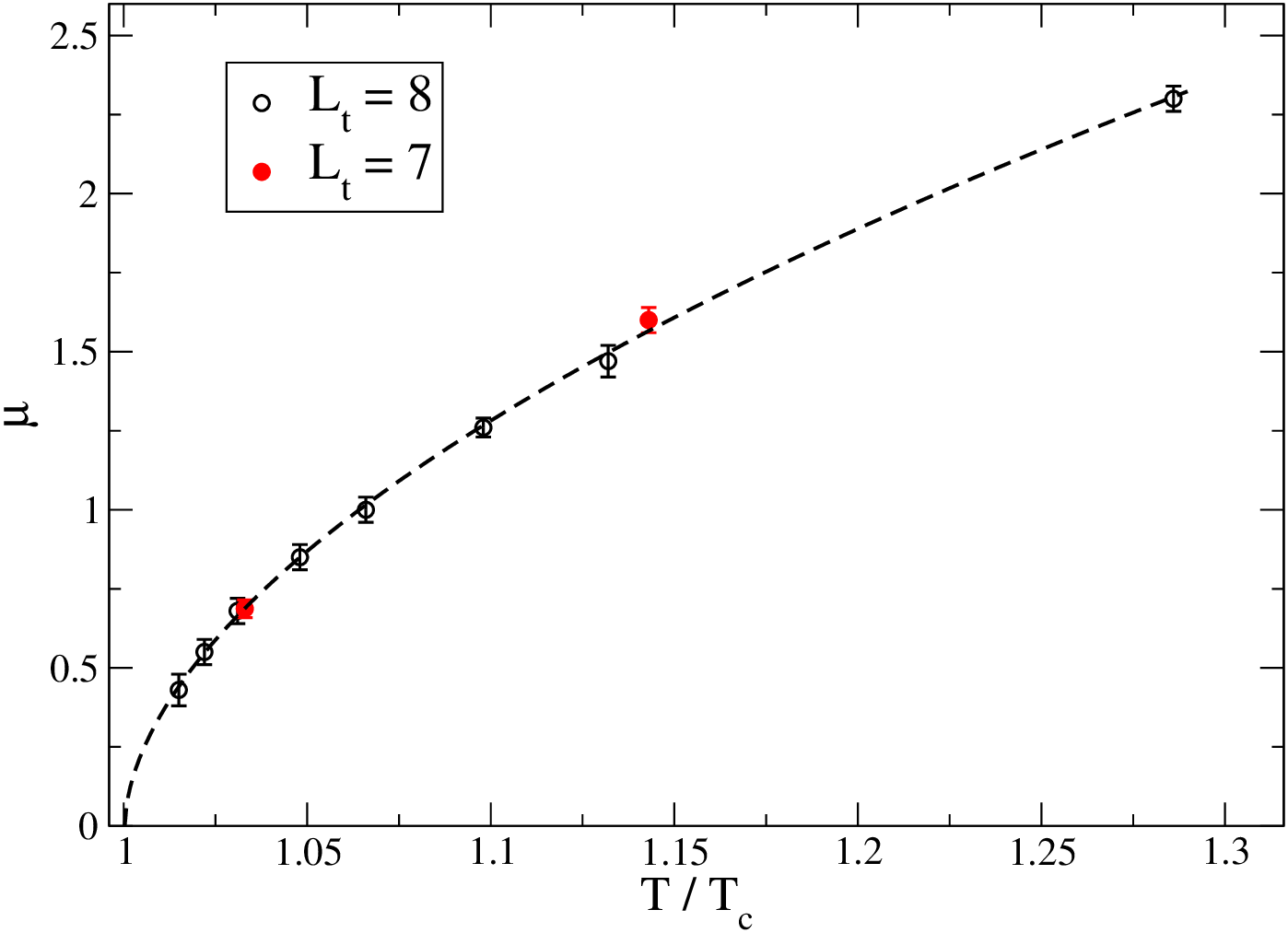}
\vspace{-0.cm}
\caption{Effective chemical potentials as a function
of $T/T_c$, obtained by a fit to Eq.~(\ref{rhok}).
The dashed line is the result of a best fit to Eq.~(\ref{critchem}).
}
\label{potchim} 
\vspace{-0.cm}
\end{center}
\end{figure}

The effective chemical potentials, obtained in this way over 
an enlarged set of temperatures above $T_c$, are displayed 
in Fig.~(\ref{potchim}). We report results from two different
lattices, $48^3 \times 8$ and $48^3 \times 7$, in order to show that
the scaling to the continuum limit holds within errors.
As a second step, following Ref.~\cite{moncon},
we have tried to fit $\hat\mu(T)$
according to a critical behavior:
\begin{equation}
\hat\mu = A\ (T - T_{\rm BEC})^{\nu'}
\label{critchem}
\end{equation}
obtaining (all data are included) $A = 4.64(20)$,
$\nu' = 0.56(3)$ and $T_{\rm BEC} = 1.0003(36)\, T_c$, with  
$\chi^2/{\rm d.o.f.} = 1.5/7$. 

We conclude that also in the $SU(3)$ pure gauge theory, if one 
interprets thermal monopole trajectories as paths describing
the quantum properties of a thermal particle ensemble, there is 
clear evidence for such ensemble to undergo BEC-like condensation
exactly at $T_c$. Notice that for $SU(3)$, since the transition 
is first order, the coincidence of the transition temperature
with $T_{\rm BEC}$ extrapolated from the critical behavior
as in Eq.~(\ref{critchem}) is not expected apriori; in particular
one could expect that the transition takes place before 
$\hat\mu$ actually goes to zero, \ie that $T_{\rm BEC} < T_c$.
However, since the $SU(3)$ transition is a weak first order 
transition, it may still be that the two temperatures coincide 
within errors. It will be interesting, in this respect, to repeat
the same analysis for $SU(N)$ gauge theories with 
$N > 3$, where the first order transition gets stronger.

\subsection{Monopole interactions}
\label{sub:monint}

The spatial distribution of thermal monopoles at a given
Euclidean time slice gives information about the mutual 
interactions of those objects. In particular, one can study
the density--density correlation function
$g_{AB}(r) \equiv \langle \varrho^A(0) \varrho^B(r) \rangle/ (\varrho^A \varrho^B)$
between any couple $A$ and $B$ of thermal monopole species,
which can be determined as the ratio between the probability  
of having a monopole of kind $B$ at distance $r$ 
from a given reference monopole of kind $A$, and the 
same probability in case the monopole locations are 
completely uncorrelated and randomly distributed, \ie
\begin{equation}
g_{AB}(r) = \frac{1}{\varrho^B} \frac{d N^B(r) }{4 \pi r^2 dr}
\end{equation}
where  $d N^B (r)$ is the number of monopoles in a spherical shell
of thickness $d r$ at distance $r$ from the reference monopole of kind $A$;
in order to minimize artifacts related to the lattice geometry,
we have used, in place of $4 \pi r^2 dr$, the actual number
of lattice sites contained  in the shell.

A value $g_{AB}(r)<1$ ($g_{AB}(r)>1$) indicates that at distance $r$ we have less (more)
particles than expected in a non-interacting medium, \ie there is a
repulsive (attractive) interaction. The determination of such correlation functions 
for $SU(2)$ has shown the existence of screened Coulomb-like attractive (repulsive) 
interactions between monopoles and antimonopoles (monopoles)~\cite{monden, born2}, with 
a strength which grows with the temperature, in agreement with expectations
based on the electric-magnetic duality~\cite{shuryak}. 

\begin{figure}[t!]
\begin{center}
\includegraphics*[width=0.65\textwidth]{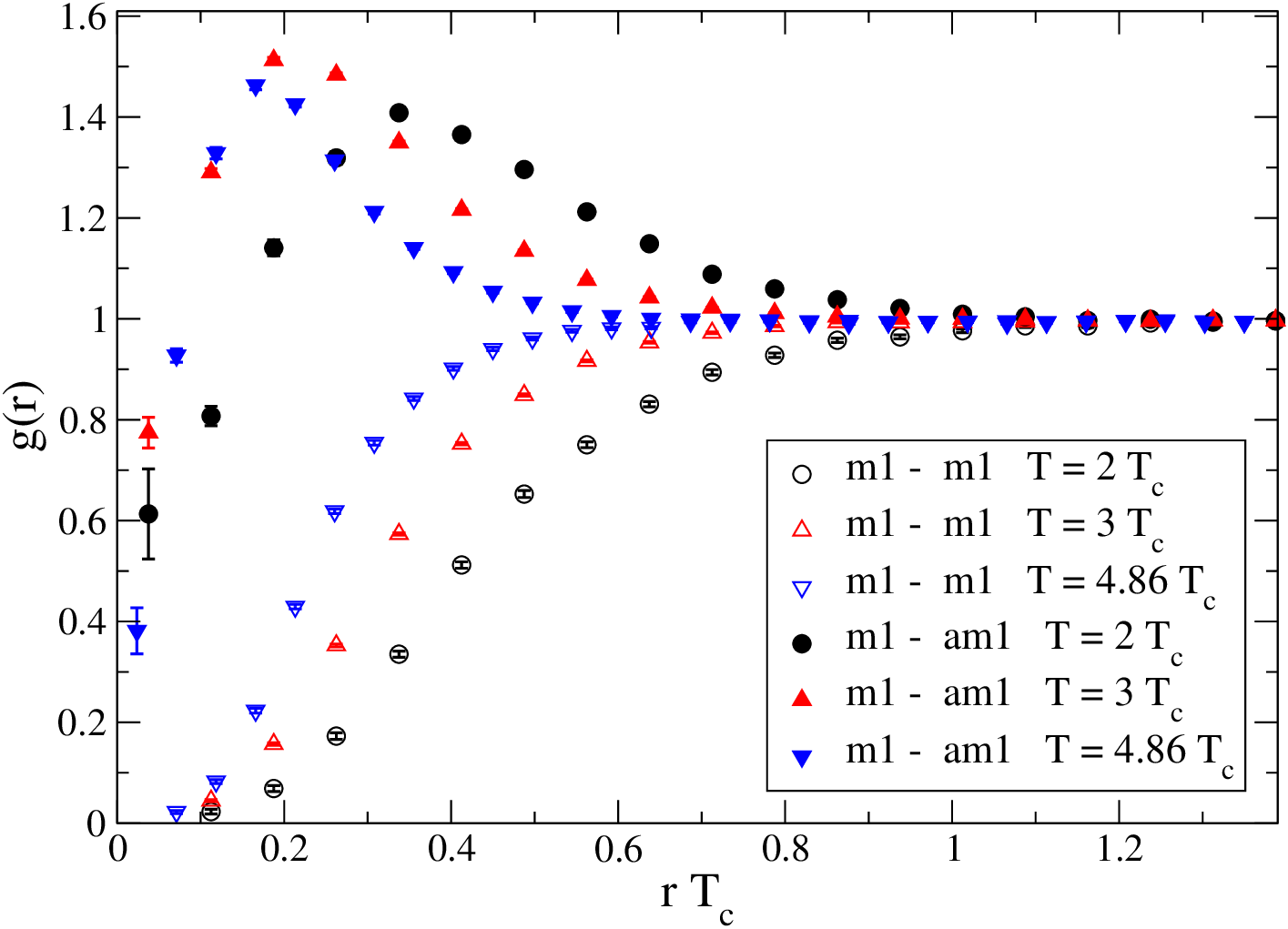}
\vspace{-0.cm}
\caption{Density correlations of monopoles (m) and antimonopoles (am) of species 1,
as a function of $r T_c$, for three different values of $T$.
}
\label{corr11} 
\vspace{-0.cm}
\end{center}
\end{figure}

For $SU(3)$, we have determined the correlation functions for three values of $T$, 
where we have collected larger statistics (up to a few thousands decorrelated configurations). 
For $T/T_c = 1.333$ we have performed simulations on two different lattice sizes and $\beta$ 
values ($48^3 \times 9$ at $\beta = 6.3368$ and $32^3 \times 6$ at $\beta = 6.0609$), in order 
to check also for continuum limit corrections, while for $T/T_c = 2,\, 3$ and $T/T_c = 4.86$ we 
have performed simulations at a single lattice spacing ($48^3 \times 6$ and $48^3 \times 4$ 
lattices at $\beta = 6.3368$ and $\beta = 6.7$). We have chosen the shell thickness to be 
$0.9$ lattice spacing in all cases.

\begin{figure}[h!]
\begin{center}
\includegraphics*[width=0.65\textwidth]{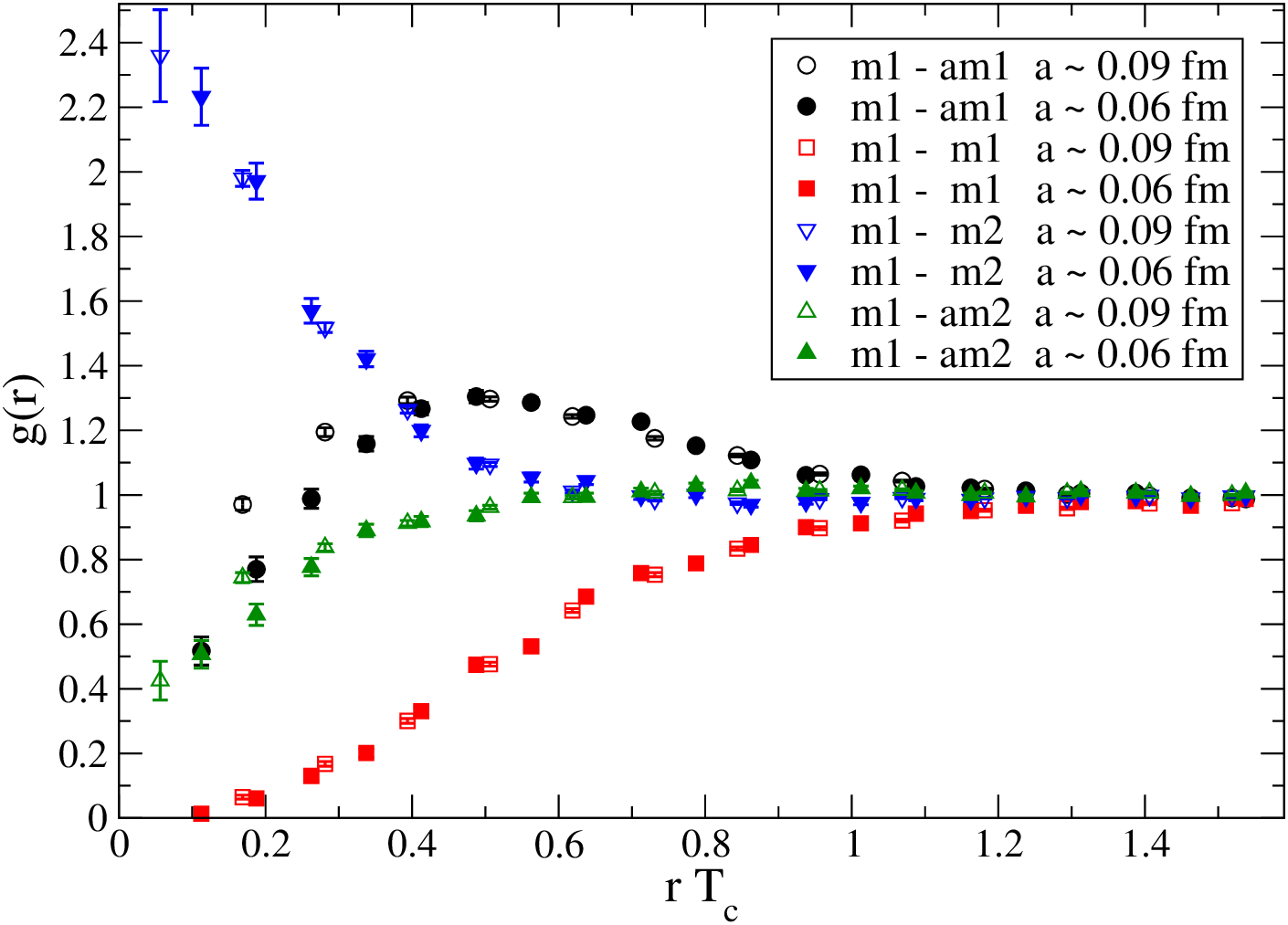}
\caption{Continuum scaling of density correlations at $T = 1.33\, T_c$.
Data have been obtained on two different lattice sizes corresponding to 
equal spatial volumes, $48^3 \times 9$ and $32^3 \times 6$.}
\label{scalinggr} 
\vspace{-0.cm}
\end{center}
\end{figure}

In Fig.~\ref{corr11} we show correlations between monopoles and antimonopoles of the same
species and for the three highest values of $T$. Data for $T/T_c = 1.333$ are reported separately 
in Fig.~\ref{scalinggr}, to better appreciate the good scaling to the continuum
limit. Results are qualitatively very similar to those obtained for $SU(2)$;
a qualitative agreement is also found with the results for monopole-monopole
interactions reported in Ref.~\cite{bornsu3}, 
which are based on the functional in Eq.~(\ref{magsu3_1}).
As for $SU(2)$ we can try to obtain information about the interaction 
potential $V(r)$ by looking
at the large distance region, where 
\begin{equation}
g_{AB}(r) \simeq \exp(-V_{AB}(r)/T) \, .
\label{yuk1}
\end{equation}
In $SU(2)$, a screened Yukawa potential,
\begin{equation}
V_{AB}(r) = \frac{\alpha_M e^{-\lambda_P r}}{r} \, ,
\label{yuk2}
\end{equation}
fits well numerical data, therefore we have tried a similar ansatz also for $SU(3)$.

An important question is whether we can describe both monopole-monopole and 
monopole-antimonopole interactions by the same (opposite) magnetic coupling $\alpha_M$,
\ie if the interaction  among monopoles and antimonopoles is Coulomb-like.
The answer is that if we try separate fits to monopole-monopole and monopole-antimonopole
correlations, different values of $\alpha_M$ are obtained, however it is possible
to perform a common fit, assuming that 
the coupling is the same, obtaining values
of $\chi^2$ which are marginally acceptable; fit results are reported in 
Table~\ref{tabalpha}. The fitted magnetic coupling $\alpha_M$ shows
a slowly increasing behavior as $T$ increases, in fair agreement with 
$SU(2)$ results and with arguments based on the electric-magnetic duality~\cite{shuryak}; values obtained for $\alpha_M$ are also in rough 
agreement with the results reported in Ref.~\cite{bornsu3}. 
The screening length shows a sizable decrease as $T$ increases.
The plasma parameter, $\Gamma \equiv \alpha_M (4 \pi \rho/ 3 T^3)^{1/3}$, stays 
always well above 1, indicating a strongly interacting behavior of the 
thermal monopole plasma.

\begin{table}
\begin{center}
\begin{tabular}{|c|c|c|c|}
\hline
$T/T_c$  &  $\alpha_M$  & $\lambda_P\, T_c$ & $\chi^2/{\rm d.o.f.}$ \\
\hline  1.333  & 3.1(4) & 0.285(11)    & 52/36       \\
\hline  2      & 4.3(5) & 0.209(6)     & 48/39       \\
\hline  3      & 5.9(5) & 0.151(3)    &  48/41      \\
\hline  4.86   & 6.4(6) & 0.114(3)    &  47/41      \\
\hline
\end{tabular}
\end{center}
\caption{Parameters of the interaction potential in Eq.~(\ref{yuk2})
obtained, for various temperatures, by a common fit to 
the monopole-monopole and monopole-antimonopole
spatial correlations of the same species.} 
\label{tabalpha}
\end{table}

\begin{figure}[h!]
\begin{center}
\includegraphics*[width=0.65\textwidth]{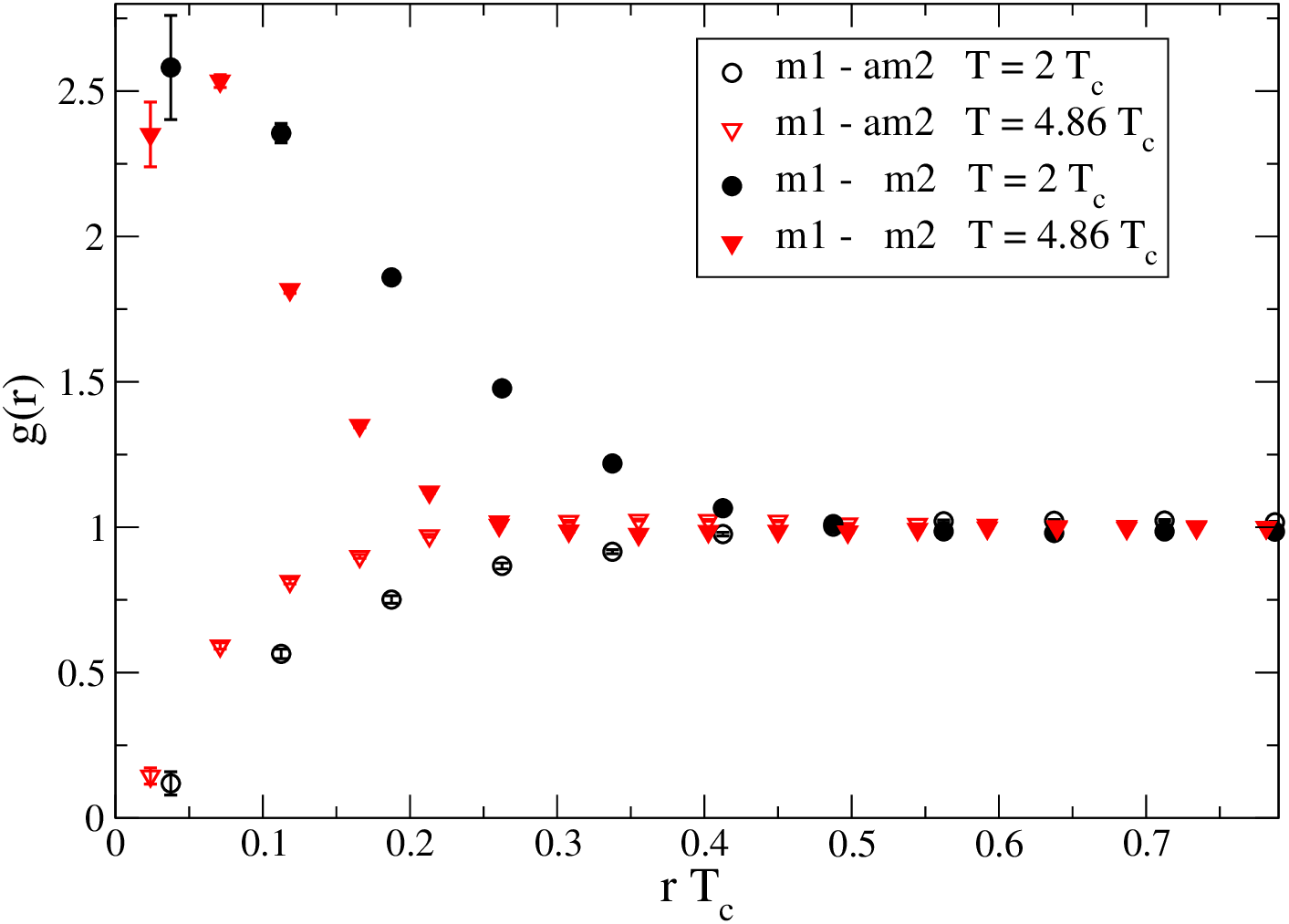}
\vspace{-0.cm}
\caption{Density correlations between monopoles and antimonopoles of different 
species (1 or 2), as a function of $r T_c$, for two values of $T$.}
\label{corr12} 
\vspace{-0.cm}
\end{center}
\end{figure}

Finally, a new, non-trivial aspect of $SU(3)$, with respect to $SU(2)$, regards 
the correlations among monopoles of different species, $g_{m^1\, m^2}(r)$ and 
$g_{m^1\, am^2}(r)$. Those are shown in Fig.~\ref{corr12} for two values of $T$. 
Correlations looks very similar to those among monopoles of the same species, 
however the sign of the interaction is opposite, with attraction between 
monopoles and repulsion between monopoles-antimonopoles of different species.
This is not surprising, if we recall that the magnetic charge operators of each
species are proportional to the corresponding roots, \ie, for $SU(3)$, to
$\lambda_3 = {\rm diag}(1,\, -1,\, 0)/2$ and $\lambda_3' = {\rm diag}(0,\, 1,\, -1)/2$,
and that the mutual, Coulomb-like interaction between monopoles of different species 
must be proportional to the scalar product of the corresponding charges~\cite{unsal}, 
\ie to ${\rm Tr} (\lambda_3' \lambda_3) = - 1/4$, while in the case of monopoles of the 
same species the corresponding product is proportional to 
${\rm Tr} (\lambda_3^2) = {\rm Tr} (\lambda_3'^2) = 1/2$.

\begin{figure}[h!]
\begin{center}
\includegraphics*[width=0.65\textwidth]{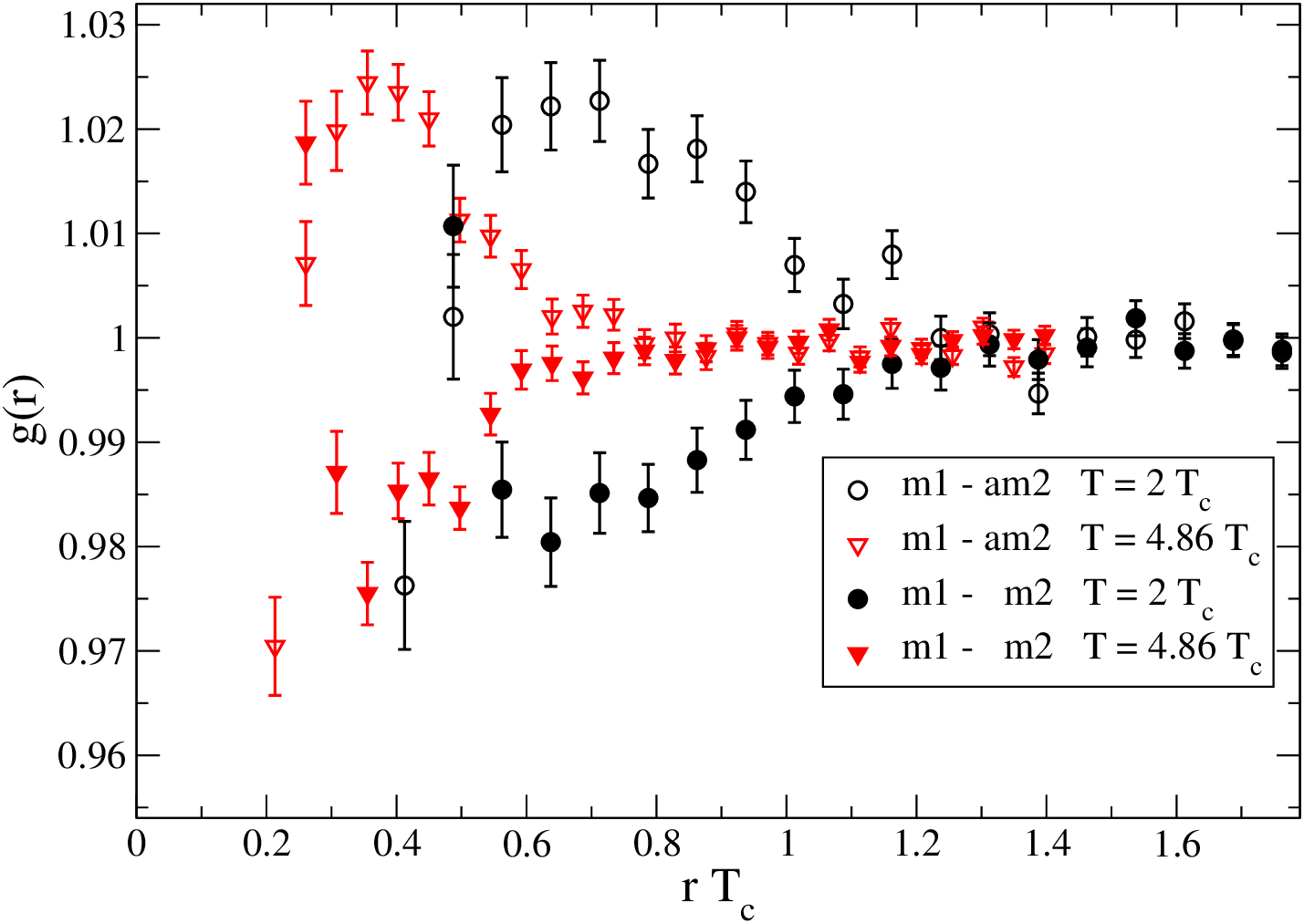}
\caption{Same as in Fig.~\ref{corr12}, with a zoom on the non-trivial 
region where the correlations between different species change sign.
}
\label{corr12zoom} 
\end{center}
\end{figure}

That suggests that, apart from the minus sign, we should be able to see also a factor
$1/2$ in the corresponding coupling. However, new, unexpected features of the correlation
functions make a fit according to Eq.~(\ref{yuk2}) unfeasible.
Indeed, a more careful observation shows that, after the first positive peak in
$g_{m^1 m^2}(r)$, a small, negative well develops for $g - 1$, corresponding 
to repulsive interaction at intermediate distances;
the opposite behavior is visible also in the monopole-antimonopole correlation, $g_{m^1 am^2}(r)$.
All that is better visible in Fig.~(\ref{corr12zoom}), where the interesting
intermediate region has been magnified.

An oscillating behavior of the density correlation function $g(r)$ is 
typical of systems with non-trivial, \eg liquid-like, properties.
A possible explanation, for instance, could be that monopoles of different species
are bound into larger objects (think, for instance, of calorons and of their
monopole constituents), and that such larger objects undergo weak repulsion,
thus explaining the inversion at intermediate distances.
It is interesting to notice that, reasoning along the same lines
of Ref.~\cite{liao_nf}, the possible formation of bound states
between monopoles of different species may explain the lower values
of $T$ which are needed to reach confinement (monopole condensation)
in $SU(3)$ with respect to $SU(2)$. All that claims for more careful 
future studies and, in particular, for an extension to larger ($N > 3$) 
gauge groups, where the pattern of interactions could be even more 
interesting and reveal fundamental aspects of Yang-Mills theories.

Correlations among different species are reported in Fig.~\ref{scalinggr} as well, 
where data for $T = 1.333\, T_c$ obtained at two different lattice spacings are compared, showing
a very good scaling to the continuum limit, apart from small deviations in the 
short distance region, \ie at the scale of the UV cutoff.

\section{Conclusions}
\label{discon}

The purpose of the present study has been that of extending 
the investigation of thermal monopoles properties to the theory 
with 3 colors. As a first step in this direction, the extension
of the Maximal Abelian projection to $SU(N)$ gauge theories 
with $N \geq 3$ has been discussed. 

We have shown that
extensions usually adopted in the literature, based on the maximization
of the diagonal components of gauge links, may not lead to a 
proper identification of the different monopole species.  
Instead, inspired by previous suggestions from the 
literature~\cite{stack_1,stack_1b,stack_2},
we have proposed and implemented an extension which 
has still a well defined Higgs field associated with it and
identifies a strict $U(1)^{(N-1)}$ residual symmetry,
leading to a proper detection of independent monopole species.

Based on that, we have presented various results regarding
the properties of thermal monopoles in the deconfined phase of
the $SU(3)$ pure gauge theory. Most properties are very
similar to those of $SU(2)$ monopoles, including the density of both monopole
species and the distribution of trajectories with multiple wrappings,
which indicate condensation of both monopole species at the 
deconfinement transition.

Spatial correlations of thermal monopoles, instead, present new, interesting
features. Correlations among monopoles of the same species 
still indicate the presence of a screened, Coulomb-like interaction,
with a magnetic coupling which increases with $T$. New interactions
however appear: monopoles of different species 
attract each other, while monopole - antimonopole pairs repel, in 
agreement with the structure of their charge operators within the 
$SU(3)$ group. Moreover, spatial correlation functions among 
monopoles of different species show a clear oscillating behavior,
with secondary long range structures, which indicate the
presence of non-trivial, \eg, liquid-like, properties,
and may be related to the formation of monopole-monopole bound states.
It is tempting, in view of that, 
to associate thermal monopoles with caloron constituents carrying 
fractional topological charges~\cite{kraan}, and to make a direct 
connection between the condensation of thermal monopoles
and the drastic change, at the phase transition, in the dependence
of the theory on the topological parameter 
$\theta$~\cite{bergman,ztn,ztn2,bdpv,dn,ztn3}.
All that claims for a further extension of the present study 
to $SU(N)$ gauge theories, with $N > 3$, where the pattern of interactions
is expected to be even more interesting, and, of course, to QCD.

\section*{Acknowledgments}
We thank A. Di Giacomo, E. Shuryak and A. Zhitnitsky for many
useful discussions. We acknowledge the use of the 
computer facilities of the 
INFN CSNIV cluster in Pisa.

\appendix

\section{Gauge Fixing Algorithm}
\label{magalg}

The algorithm for the maximization of the functional in Eq.~(\ref{magsu3_2}) 
follows closely that commonly used in the $SU(2)$ case, \ie a combination of 
local maximization-overrelaxation, which we briefly review in the following.

If we perform a gauge transformation which is non-trivial only in 
one lattice site $n$, $G(n)$, then the part of the functional 
in Eq.~(\ref{magsu2}) which depends on $G(n)$ is 
\begin{equation}
{\rm Tr} \left( G^\dagger(n) \sigma^3 G(n) X (n) \right) \, ,
\label{localmagsu2}
\end{equation}
where $X(n)$ is defined in Eq.~(\ref{Xdef}). It is easy to find the $SU(2)$ element 
maximizing the expression~(\ref{localmagsu2}). $G(n)$, as any $SU(2)$ element, can 
always be written in the form 
\begin{equation*}
G(n) = (g_0 {\rm Id} + i g_1 \sigma_1 + i g_2 \sigma_2) 
( \sqrt{1 - g_3^2}\ {\rm Id} + i g_3 \sigma_3 )
\end{equation*}
and the expression in Eq.~(\ref{localmagsu2}) is independent of $g_3$. 
Hence, without loss of generality, we can parameterize $G(n)$ as follows
\begin{equation}
G(n) = \cos \alpha\ {\rm Id} + i \sin \alpha\
(\cos \phi \sigma_1 + \sin \phi \sigma_2) \, ,
\label{Gparam}
\end{equation}
so that
\begin{equation*}
G^\dagger(n) \sigma^3 G(n) = \cos (2 \alpha)\ \sigma_3 + \sin (2 \alpha) 
(\sin \phi\ \sigma_1 - \cos \phi\ \sigma_2) \, .
\end{equation*}
$X(n)$ is a traceless adjoint operator, \ie one can write 
$X = \vec x \cdot \vec \sigma$, hence
\begin{equation}
{\rm Tr} \left( G^\dagger(n) \sigma_3 G(n) X (n) \right) =
2 x_3 \cos (2\alpha) + 2 (x_1 \sin \phi  - x_2 \cos \phi) (\sin 2 \alpha) \, .
\label{localmax}
\end{equation}
Last expression has a maximum when the unit vector 
$(\sin(2\alpha) \cos \phi,  \sin(2\alpha) \sin \phi,  \cos(2\alpha))$
is parallel to $(-x_2,x_1,x_3)$, hence
\begin{equation}
\begin{aligned}
\cos \phi_{\rm max} &= \frac{- x_2}{\sqrt{x_1^2 + x_2^2}}\, ;  \\
\sin \phi_{\rm max} &= \frac{  x_1}{\sqrt{x_1^2 + x_2^2}}\, ; \\
\cos (2 \alpha_{\rm max}) &= \frac{x_3}{\sqrt{x_1^2 + x_2^2 + x_3^2}} \, . 
\label{su2sol}
\end{aligned}
\end{equation}
The gauge-fixing algorithm consists of sweeps of one-site gauge transformations 
which locally maximize the MAG functional; the algorithm is stopped when the 
adjoint operator $X(n)$ turns out to be diagonal within a given precision, 
\ie when the average squared modulus of the non-diagonal contribution,  
$\sum_n (x_1(n)^2 + x_2(n)^2)/V$, where $V$ is the lattice volume, 
goes below a given threshold. The algorithm can be accelerated by 
using overrelaxation~\cite{creutz,mandula-ogilvie}, \ie by choosing 
\begin{equation}
G(n) = (G_{\rm max}(n))^\omega =
\cos (\omega \alpha_{\rm max})\ {\rm Id} + i \sin (\omega \alpha_{\rm max})\ 
(\cos \phi_{\rm max}\ \sigma_1 + \sin \phi_{\rm max}\ \sigma_2) \, ;
\label{overrelax_su2}
\end{equation}
a value $\omega \simeq 1.8$ reveals to be optimal.

Let us now switch to the general $SU(N)$ functional defined in 
Eq.~(\ref{magsu3_2}). In this case, inspired by the Cabibbo-Marinari algorithm 
for heat-bath updating in $SU(N)$~\cite{cabibbo}, we will adopt a procedure of 
local maximization over $SU(2)$ subgroups. Let us consider again a gauge 
transformation which is non-trivial only at site $n$ and suppose further that 
$G(n)$ is non-trivial only in the $SU(2)$ subgroup corresponding to rows and 
colums $i$ and $j$ ($i < j$). The functional to be maximized by $G(n)$ is 
\begin{equation}
{\rm Tr} \left( G^\dagger(n) \tla G(n) \tilde X (n) \right) \, .
\end{equation}
Also in this case one proves that, if the $SU(2)$ expression of $G(n)$ is
$(g_0 {\rm Id} + i g_1 \sigma_1 + i g_2 \sigma_2) 
( \sqrt{1 - g_3^2}\ {\rm Id} + i g_3 \sigma_3 )$, 
then $g_3$ is irrelevant and one can ignore it. 
Therefore we fix 
\begin{equation*}
G(n)|_{\rm SU(2)} =  \cos \alpha\ {\rm Id} + i \sin \alpha\
(\cos \phi \sigma_1 + \sin \phi \sigma_2)  \, .
\end{equation*}
We have that $(G^\dagger(n) \tla G(n))_{ij} = \delta_{ij} \tla_i$ 
for $i$ or $j$ outside the given $SU(2)$ subgroup, while for both $i$ and $j$ 
inside it has the form:
\begin{equation}
G^\dagger(n) \tla G(n)|_{\rm SU(2)} = 
\frac{\tla_i + \tla_j}{2}\ {\rm Id}
+ \frac{\tla_i - \tla_j}{2} \left(
\cos (2 \alpha)\ \sigma_3 + 
\sin (2 \alpha) 
(\sin \phi\ \sigma_1 - \cos \phi\ \sigma_2) \right) \, .
\end{equation}
Regarding $\tilde X(n)$, we can always parameterize its restriction
to the $SU(2)$ subgroup as follows
\begin{equation}
\left(\begin{matrix}
\tilde X(n)_{ii} & \tilde X(n)_{ij} \cr 
\tilde X(n)_{ji} & \tilde X(n)_{jj} \cr
\end{matrix}
\right) \equiv  \frac{\tilde X(n)_{ii} + \tilde X(n)_{jj}}{2}\ {\rm Id}
 + \vec x \cdot \vec \sigma
\end{equation}
hence
\begin{equation}
{\rm Tr} \left( G^\dagger(n) \tla G(n) \tilde X (n) \right) =
C + (\tla_i - \tla_j) \left[
x_3 \cos (2\alpha) + 2 (x_1 \sin \phi  - x_2 \cos \phi) (\sin 2 \alpha) \right]
\, ,
\label{localmax_su3}
\end{equation}
where $C$ is a constant. 
If the $\tla$ eigenvalues are ordered, as we have assumed in Section~\ref{2C}, 
then $\tla_i - \tla_j > 0$ and the solution for the local maximum over the chosen
subgroup is obtained exactly as in Eq.~(\ref{su2sol}), otherwise a global minus 
sign would apply, $(x_1, x_2, x_3) \to (-x_1, -x_2, -x_3)$.
Overrelaxation also proceeds as in Eq.~(\ref{overrelax_su2}): we have found that 
a coefficient similar to that used for $SU(2)$, $\omega \sim 1.8$, is optimal 
also in the $SU(3)$ case. 
The local overrelaxation is repeated iteratively over all possible $N\, (N-1)/2$ 
subgroups and over all lattice sites. The algorithm is stopped when the average 
of the squared moduli of the non-diagonal elements of $\tilde X(n)$ goes below 
a given threshold, which has been set to $10^{-10}$.

The maximization of the functional in Eq.~(\ref{magsu3_1}) can be performed
locally over the $SU(2)$ subgroups as well. Let us consider a gauge transformation 
$G(n)$ belonging to the first subgroup, $i,j = 1$ and 2, at first: in this case, 
since the term ${\rm Tr} \left( U_\mu(n) \lambda_8 U^{\dagger}_\mu(n) \, \lambda_8\right)$
is invariant under such gauge transformations, everything goes 
exactly as specified previously with $\tla = \lambda_3$. 
On the other hand the functional in Eq.~(\ref{magsu3_1}) is symmetric
over the $SU(2)$ subgroups, hence one can use exactly the same procedure for all 
subgroups. As we have stressed above, no diagonal 
local adjoint operator is naturally associated with
this version of the maximal abelian gauge. Anyway one can prove that, when a 
maximum of the expression (\ref{magsu3_1}) is reached, each of the operators
$X(n)$, $X'(n)$ and $X''(n)$, defined as in Eq.~(\ref{Xdef}) with 
$\tilde \lambda = \lambda_3, \lambda_3'$ and $\lambda_3''$ respectively,
is diagonal when restricted to the corresponding subgroup, \ie one has
\begin{eqnarray}
|X_{12}(n)|^2 = |X'_{23}(n)|^2 = |X''_{13}(n)|^2 = 0
\end{eqnarray}
for each site $n$ (see, \eg, Ref.~\cite{stack_1}). Such condition can be taken as a stopping 
criterion in this case.

\section{Extraction of the diagonal part of gauge links}
\label{takediag}

For the  $SU(2)$ gauge theory, the extraction of the Abelian phases, \ie 
the degrees of freedom related to the residual $U(1)$ invariance gauge group, 
coincides with taking the phases of the diagonal elements of the gauge links.

In $SU(N)$ the procedure is less trivial. Indeed, writing the diagonal element of 
the generic $SU(N)$ matrix $U$ in the form $U_{ii} = |U_{ii}| \exp{(i \varphi_i})$, 
it is not guaranteed that ${\rm diag} (\exp{(i \varphi_1)}, \exp{(i \varphi_2)}, \dots, 
\exp{(i \varphi_N}))$ belongs to $SU(N)$, since, in general, 
${\rm mod} (\sum_i \varphi_i,2\pi) = \delta \varphi$ with $\delta \varphi \neq 0$.

One simple procedure, adopted in some $SU(3)$ studies~\cite{mag3_3}, 
is to define an $SU(N)$ diagonal element 
$u = {\rm diag} (\exp{(i \phi_1)}, \exp{(i \phi_2)}, \dots , \exp{(i \phi_N)})$
with $\phi_i = \varphi_i - \delta \varphi / N$ and extracting the Abelian phases 
from it. A more accurate procedure~\cite{stack_1} is instead to project the original 
$SU(N)$ element $U$ onto the closest diagonal element $u$ by maximizing 
\begin{equation}
{\rm Re}( {\rm tr} (u U^\dagger)) = \sum_i |U_{ii}| \cos (\phi_i - \varphi_i)\, .
\end{equation}
Here we present an approximate but quite simple and accurate procedure to get 
the maximum. It is easy the check that, due to the constraint 
${\rm mod} (\sum_i \phi_i,2\pi) = 0$, the condition for stationary points of 
${\rm Re}( {\rm tr} (u U^\dagger))$ is to have
\begin{equation}
{|U_{11}|}\ {\sin (\phi_1 - \varphi_1)} = 
{|U_{22}|}\ {\sin (\phi_2 - \varphi_2)} = 
\dots = 
{|U_{NN}|}\ {\sin (\phi_N - \varphi_N)} \, .
\label{stationary}
\end{equation}
On the other hand, for links of gauge fixed configurations, usually
$\delta \varphi$ is a small quantity and on the maximum we expect
$|\varphi_i - \phi_i| \ll 1$, so that we can approximate 
$\sin (\phi_i - \varphi_i) \sim (\phi_i - \varphi_i)$ and obtain
the following solution to Eq.~(\ref{stationary})
\begin{eqnarray}
\phi_i = \varphi_i - \delta \varphi 
\frac{|U_{ii}|^{-1}}{\sum_j |U_{jj}|^{-1}} \, .
\label{solution_stationary}
\end{eqnarray}
The extraction of the Abelian phases $\theta^k$ then proceeds as 
specified in Eq.~(\ref{thetadef}).

\section{Dependence of results on the choice of $\tla$}
\label{tlachange}

Even if the choice made in Eq.~(\ref{fixtla}), \ie of treating all monopole
species symmetrically, is the most natural, it is interesting to understand 
what are the effects of a different choice of $\tla$.

\begin{figure}[h!]
\begin{center}
\includegraphics*[width=0.65\textwidth]{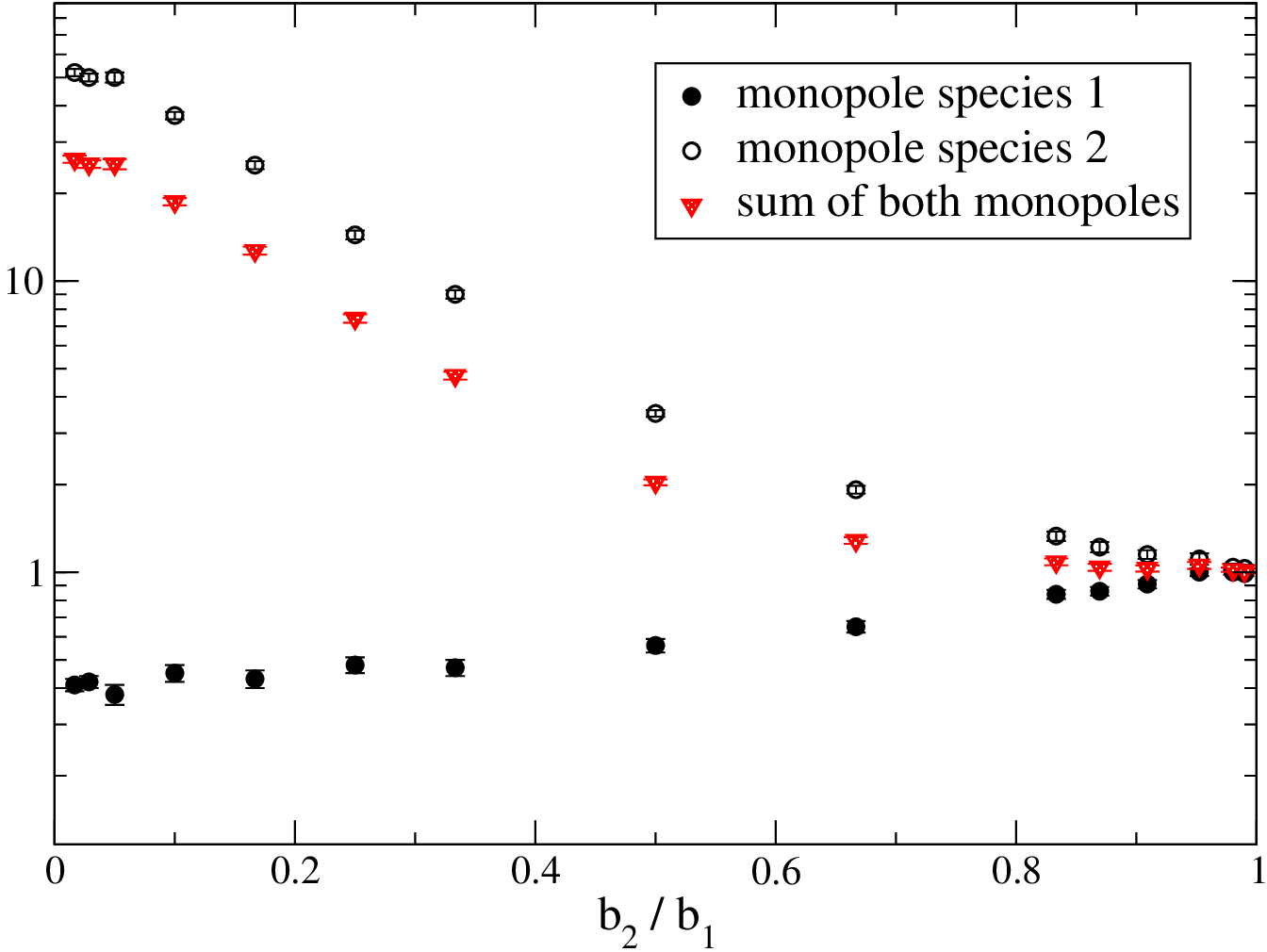}
\vspace{-0.cm}
\caption{Density of monopole currents obtained, for different choices of $\tilde \lambda$,  on a $16^4$ lattice 
at $\beta = 6.3$,
normalized to the density obtained for the 
symmetric choice of $\tilde \lambda$ ($b_1 = b_2$) 
adopted in this study.}
\label{dens_lambdatilde} 
\vspace{-0.cm}
\end{center}
\end{figure}

For that reason, in Fig.~(\ref{dens_lambdatilde}) we show the total density 
of 3D cubes containing monopole currents, separately for each monopole
species, determined on a $16^4$ lattice at $\beta = 6.3$ and
as a function of $b_2/b_1$ (see Eq.~(\ref{tla1})).
Densities are normalized to that obtained at $b_2/b_1 = 1$, which corresponds 
to our original choice. We report also the average of the two densities.
Results show that, for small variations of $b_2/b_1$ around 1, the densities of the 
two species change but their sum remains stable. However, when $b_2 \ll b_1$, the density
for species 2 grows by more than one order of magnitude: that can be related to the 
appearance of significant lattice artifacts in the identification of that kind of 
monopoles, due to the fact that $b_2 \sim 0$, as discussed in Section~\ref{2C}.
Qualitatively similar results are obtained when one looks at thermal monopoles.

\begin{figure}[h!]
\begin{center}
\includegraphics*[width=0.65\textwidth]{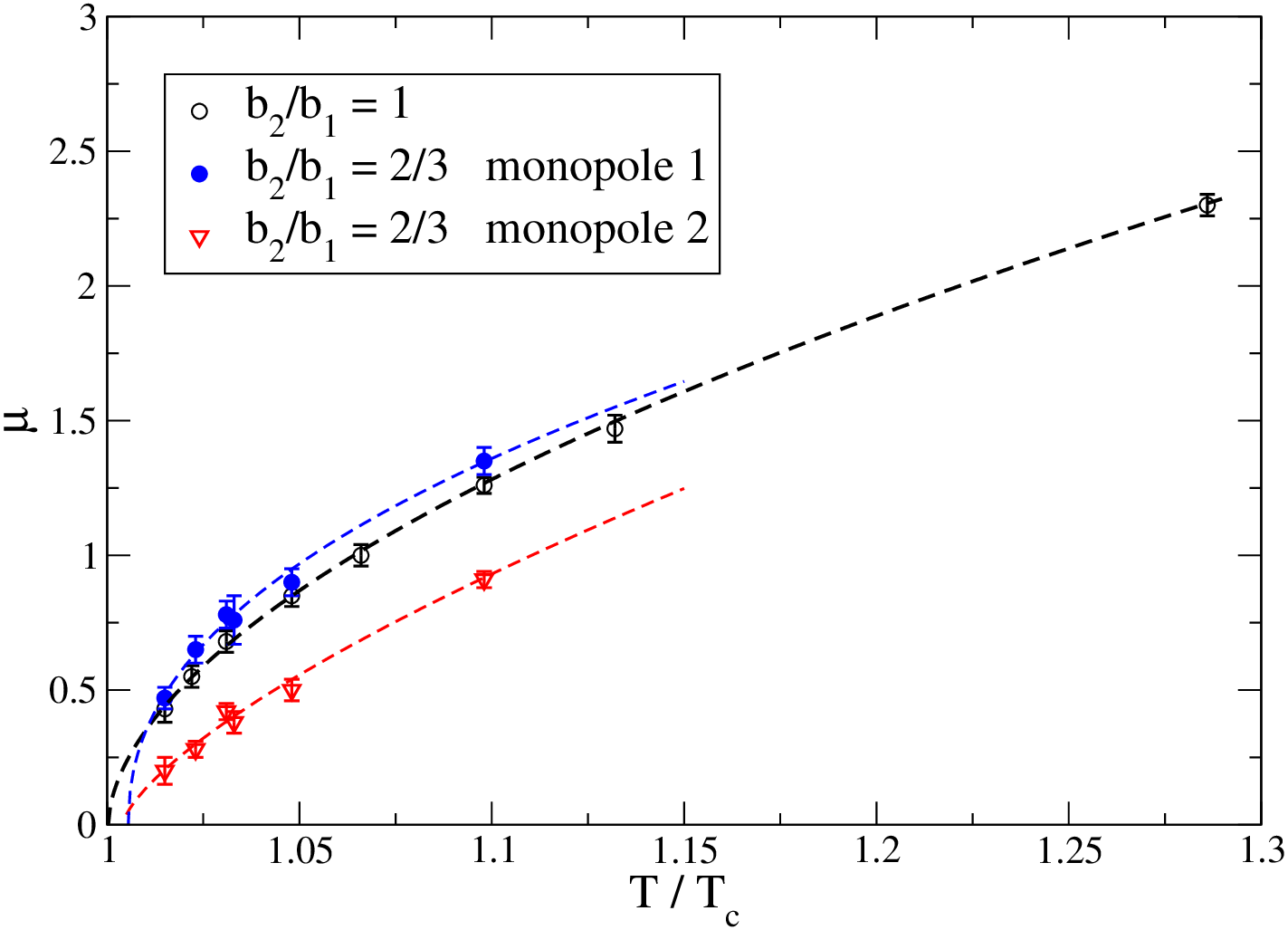}
\vspace{-0.cm}
\caption{Effective chemical potentials as a function
of $T/T_c$, obtained by a fit to Eq.~(\ref{rhok}), for different
choices of $\tla$ and for the different thermal monopole species.
The dashed lines are the result of a best fit to Eq.~(\ref{critchem})
(see text).
}
\label{fitmulti} 
\vspace{-0.cm}
\end{center}
\end{figure}

It is interesting to investigate which features remain stable when the 
ratio $b_2/b_1$ is modified. In Fig.~(\ref{fitmulti}) we report the analogous
of Fig.~(\ref{potchim}), including chemical potentials obtained for the different
monopole species and for $b_2/b_1 = 2/3$. The effective chemical potentials show 
a clear dependence on $\tla$. However, when one tries to fit data according
to Eq.~(\ref{critchem}), to infer the temperature at which each monopole
species shows signals of condensation, one finds
$T_{\rm BEC} = 1.005(4)\, T_c$, with  
$\chi^2/{\rm d.o.f.} = 1.5/3$, for monopole species 1, and 
$T_{\rm BEC} = 1.004(5)\, T_c$, with  
$\chi^2/{\rm d.o.f.} = 3/3$, for monopole species 2. Hence, we infer
that the temperature at which condensation seems to happen is stable
under relatively small variations of $b_2/b_1$. This is in nice
agreement with the idea put forward in \cite{nabianchi, bdd} that
the detection of monopoles is abelian-projection dependent but monopole condensation
it is not.

\end{document}